\documentclass[twocolumn, tighten]{aastex631}

\usepackage{amsmath,amsopn,amsxtra,txfonts}
\usepackage{comment} 
\usepackage{quotes}
\usepackage{natbib} 
\usepackage{microtype}
\usepackage{hyperref}
\usepackage{colortbl}
\hypersetup{
colorlinks = true,
linkcolor = cyan,
citecolor = cyan,
filecolor = cyan,
urlcolor = cyan}
\newcommand{\threedhst}{\hbox{3D-HST}}

\newcommand{\editone}[1]{\textcolor{black}{#1}}
\newcommand{\edittwo}[1]{\textcolor{black}{#1}}

\newcommand{\prospector}{\texttt{Prospector}}
\newcommand{\pa}{\texttt{Prospector-$\alpha$}}
\newcommand{\ugi}{\emph{ugi}}
\newcommand{\sugi}{(\emph{ugi})\hbox{$_s$}}
\newcommand{\uvj}{\emph{UVJ}}
\newcommand{\su}{\emph{u$_s$}}
\newcommand{\sg}{\emph{g$_s$}}
\newcommand{\si}{\emph{i$_s$}}
\newcommand{\eazy}{\texttt{EAZY}}
\newcommand{\eazypy}{\texttt{eazy-py}}
\newcommand{\fsps}{\texttt{FSPS}}

\newcommand{\hb}{\hbox{H$\beta$}}
\newcommand{\oiii}{\hbox{[\ion{O}{3}]}}
\newcommand{\oii}{\hbox{[\ion{O}{2}]}}
\newcommand{\nii}{\hbox{[\ion{N}{2}]}}

\newcommand{\jwst}{\textit{JWST}}
\newcommand{\hst}{\textit{HST}}
\newcommand{\herschel}{\textit{Herschel}}
\newcommand{\spitzer}{\textit{Spitzer}}

\submitjournal{\apj}

\begin{document}
\definecolor{aggiemaroon}{HTML}{500000}

\title{Beyond UVJ: Color Selection of Galaxies in the JWST Era}
 \shorttitle{Beyond UVJ: Color Selection of Galaxies in the JWST Era}
 \shortauthors{Antwi-Danso et al.} 

 \author[0000-0002-0243-6575]{Jacqueline Antwi-Danso}
\affiliation{George P. and Cynthia Woods Mitchell Institute for Fundamental Physics and Astronomy, Texas A\&M University, College Station, TX, USA}
\affil{Department of Physics and Astronomy, Texas A\&M University, 4242 TAMU, College Station, TX, USA}
\author[0000-0001-7503-8482]{Casey Papovich}
\affil{George P. and Cynthia Woods Mitchell Institute for Fundamental Physics and Astronomy, Texas A\&M University, College Station, TX, USA}
\affil{Department of Physics and Astronomy, Texas A\&M University, 4242 TAMU, College Station, TX, USA}
\author[0000-0001-6755-1315]{Joel Leja}
\affiliation{Department of Astronomy \& Astrophysics, The Pennsylvania State University, University Park, PA 16802, USA}
\affil{Institute for Computational \& Data Sciences, The Pennsylvania State University, University Park, PA, USA}
\affil{Institute for Gravitation and the Cosmos, The Pennsylvania State University, University Park, PA 16802, USA}
\author[0000-0001-9002-3502]{Danilo Marchesini}
\affiliation{Department of Physics and Astronomy, Tufts University, Medford, MA, USA}
\author[0000-0002-7248-1566]{Z. Cemile Marsan}
\affiliation{Department of Physics and Astronomy, York University, Toronto, Ontario, Canada}
\author[0000-0003-3243-9969]{Nicholas S. Martis}
\affiliation{Department of Astronomy and Physics, St. Mary's University, Halifax, Nova Scotia, Canada}
\affil{National Research Council of Canada, Herzberg Astronomy \& Astrophysics Research Centre, 5071 West Saanich Road, Victoria, BC, Canada, V9E 2E7}
\author[0000-0002-2057-5376]{Ivo Labbé}
\affiliation{Center for Astrophysics \& Supercomputing, Swinburne University, Hawthorn, Victoria, Australia}
\author[0000-0002-9330-9108]{Adam Muzzin}
\affiliation{Department of Physics and Astronomy, York University, Toronto, Ontario, Canada}
\author[0000-0002-3254-9044]{Karl Glazebrook}
\affiliation{Center for Astrophysics \& Supercomputing, Swinburne University, Hawthorn, Victoria, Australia}
\author[0000-0001-5937-4590]{Caroline M. S. Straatman}
\affiliation{Department of Physics and Astronomy, Ghent University, Krijgslaan 281 S9, B-9000 Gent, Belgium}
\affil{Max-Planck Institut für Astronomie, Königstuhl 17, D-69117, Heidelberg, Germany}
\author[0000-0001-9208-2143]{Kim-Vy H. Tran}
\affiliation{School of Physics, University of New South Wales, Kensington, Australia}
\affiliation{ARC Centre for Excellence in All-Sky Astrophysics in 3D}

\correspondingauthor{Jacqueline Antwi-Danso}
\email{jadanso@tamu.edu}
\begin{abstract}
\noindent We present a new rest-frame color-color selection method using ``synthetic $u_s-g_s$ and $g_s-i_s$'', \sugi\ colors to identify star-forming and quiescent galaxies.  Our method is similar to the widely-used $U-V$ versus $V-J$ ($UVJ$) diagram.  However, \uvj\ suffers known systematics. Spectroscopic campaigns have shown that \uvj-selected quiescent samples at $z \gtrsim 3$ include $\sim 10-30\%$ contamination from galaxies with dust-obscured star formation and strong emission lines.  Moreover, at $z>3$, $UVJ$ colors are extrapolated because the rest-frame J-band shifts beyond the coverage of the deepest bandpasses at $< $ 5~\micron\ (typically \spitzer/IRAC 4.5 $\mu m$ or future \jwst/NIRCam observations).  We demonstrate that  \sugi\ offers improvements to \uvj\ at $z>3$, and  can be applied to galaxies in the \jwst\ era.  We apply \sugi\ selection to galaxies at $0.5<z<6$ from the (observed) 3D-HST and UltraVISTA catalogs, and to the (simulated) JAGUAR catalogs. We show that extrapolation can affect $(V-J)_0$ color by up to 1 magnitude, but changes $(\sg-\si)_0$ color by $\leq$ 0.2~mag,  even at $z\simeq 6$.  While \sugi--selected quiescent samples are comparable to $UVJ$ in completeness (both achieve $\sim$85-90\% at $z=3-3.5$), $\sugi$ reduces contamination in quiescent samples by nearly a factor of two, from $\simeq$35\% to $\simeq$17\% at $z=3$, and from $\simeq $60\% to $\simeq $33\% at $z=6$.  This leads to improvements in the true-to-false-positive ratio (TP/FP),  where we find TP/FP $\gtrsim$ 2.2  for \sugi\ at $z \simeq 3.5 - 6$, compared to TP/FP $<$ 1 for \uvj-selected samples. This indicates that contaminants will outnumber true quiescent galaxies in \uvj\ at these redshifts, while \sugi\ will provide higher-fidelity samples.    
\end{abstract}
\section{Introduction}\label{section:intro}
Over the past two decades, deep extragalactic surveys have yielded impressive numbers of galaxies in the high redshift universe, thereby improving our understanding of galaxy evolution. The development of sensitive near-infrared (NIR) imagers such as Magellan/FourStar \citep{persson2013}, VLT/HAWK-I \citep{kisslerpatig2008}, ESO/VISTA \citep{sutherland2015}, and UKIRT/WFCAM \citep{casali2007} and their incorporation into multiwavelength surveys have yielded thousands of galaxy candidates at z = 3 -- 6.  The addition of medium--band filters \editone{with deep NIR surveys such as NMBS \citep{whitaker2011} and ZFOURGE \citep{straatman2016}} have additionally improved the color selection of galaxies at these redshifts, enabling the detection and characterization of red galaxies over cosmic time (e.g. \citealt{marchesini2010}; \citealt{straatman2014}, \citealt{spitler2014}; \citealt{patel2017}). \editone{Using these data, we have demonstrated} that we can accurately determine redshifts and distinguish between star-forming and quiescent galaxies with photometric data alone. Similarly, surveys of deep fields and lensing clusters using the {\emph {Hubble Space Telescope}} (\emph{HST}) have revealed hundreds of galaxies and galaxy candidates at $z = 7-10$, enabling us to identify and characterize some of the very first sources in the Universe (\citealt{grogin2011}, \citealt{koekemoer2011}, \citealt{bradley2012, bradley2014}, \citealt{ellis2013}, \citealt{schmidt2014}, \citealt{lotz2017}, \citealt{salmon2018}, \citealt{coe2019}). 

The recently-launched {\emph{James Webb Space Telescope}} (\jwst) has the potential to do the same, detecting galaxies well out to $z \sim$ 12 (\citealt{behroozi2020}, \citealt{robertson2022}) due to its unparalleled NIR imaging and spectroscopic capabilities. The Near-InfraRed Camera (NIRCam) and the Mid--InfraRed Instrument (MIRI) on \jwst\ are designed for broadband photometry from 0.7 $-$ 25.5 microns with unprecedented resolution and sensitivity. Accepted ERS, GTO, and GO\footnote{Early Release Science, Guaranteed Time Observations, and General Observer} programs will carry out the first observations with \jwst, test each instrument’s capabilities, and set the tone for future science with Cycle 1 and beyond. These will allow us to extend observations of the galaxy rest-frame UV luminosity function to higher redshifts ($z >$ 10) and fainter luminosity as well as detect rest-frame optical and NIR spectral energy distributions (SEDs) of high redshift galaxies. Even with these amazing capabilities, spectroscopy of a large number of sources with \jwst\ is unfeasible due to its short operational lifespan, and as MIRI has the smallest field of view of all the imaging instruments, we will obtain mid-IR data for less than a third of the survey area of our largest upcoming programs. 
As a result, most of the early surveys  will carry out broadband imaging spanning $\sim$0.9-4.4 microns, with the NIR observations being instrumental for characterizing the rest-frame optical of z $\sim$7 galaxies, similar to what was done with \spitzer/IRAC photometry (e.g., \citealt{labbe2013}, \citealt{oesch2014}, \citealt{smit2015}, \citealt{roberts-borsani2016}, \citealt{castellano2017}). 

For this reason, it is incumbent that we come up with an efficient and accurate method to select galaxies based on photometric colors alone, a method that retains its efficiency and accuracy in the \jwst\ era. The most widely-used method for $z\leq4$ galaxies is the \uvj\ diagram. Star-forming and quiescent galaxies have been shown to exhibit a bimodality in the rest-frame $U-V$ versus $V-J$ color-color space out to $z \sim$ 3 (\citealt{labbe2005}, \citealt{williams2009}, \citealt{muzzin2013}, \citealt{whitaker2011}, \citealt{straatman2014, straatman2016}). This bimodality correlates with specific star formation rate, MIPS 24 micron flux down to implied SFRs of $\sim 40 M_{\odot} yr^{-1}$ (\citealt{brammer2011}), and morphology (\citealt{papovich2012,papovich2015,patel2012}). \editone{Additionally, dusty star-forming and quiescent galaxies are difficult to distinguish in single-color diagrams (e.g. $U$-$V$ alone; \citealt{wolf2009}, \citealt{brammer2009}, \citealt{ leja&tacchella&conroy2019}), as both SED types exhibit red colors and peak at 1 $\mu$m in the rest-frame. The physical motivation then for including a second color (e.g. $V$-$J$) is to add a longer wavelength baseline in order to trace out the rest-frame NIR, where dusty star-forming galaxies tend to have redder colors and quiescent galaxies have bluer colors.} Consequently, the \uvj\ diagram is a useful and efficient way of selecting different galaxy populations in large photometric datasets. 

Despite its utility, the \uvj\ diagram is fraught with problems at $z >$3. Due to the growing number of spectroscopically-confirmed quiescent samples at these redshifts, we now know that \uvj-selected quiescent samples have 21-30\% contamination from galaxies with significant levels of ongoing star-formation (e.g. \citealt{schreiber2018}, \citealt{forrest2020_survey}). There are several reasons for this.  One is that the \hb+\oiii\ lines can boost the $V$-band flux, causing galaxies with strong emission lines to appear redder in $U-V$ than they actually are. Additionally, $J$-band fluxes at $z>3$ are less accurate because they require extrapolation. This occurs because the $J$-band corresponds to the \spitzer/IRAC bands at these redshifts, for which \editone{the data is often shallow (typically by $\simeq$2.5 mag, if it is available at all (\citealt{Sanders2007}, \citealt{Ashby2015})}. 

\begin{figure*}
\begin{center}
\includegraphics[trim={0cm 4cm 0cm 3cm},clip, scale=0.48]{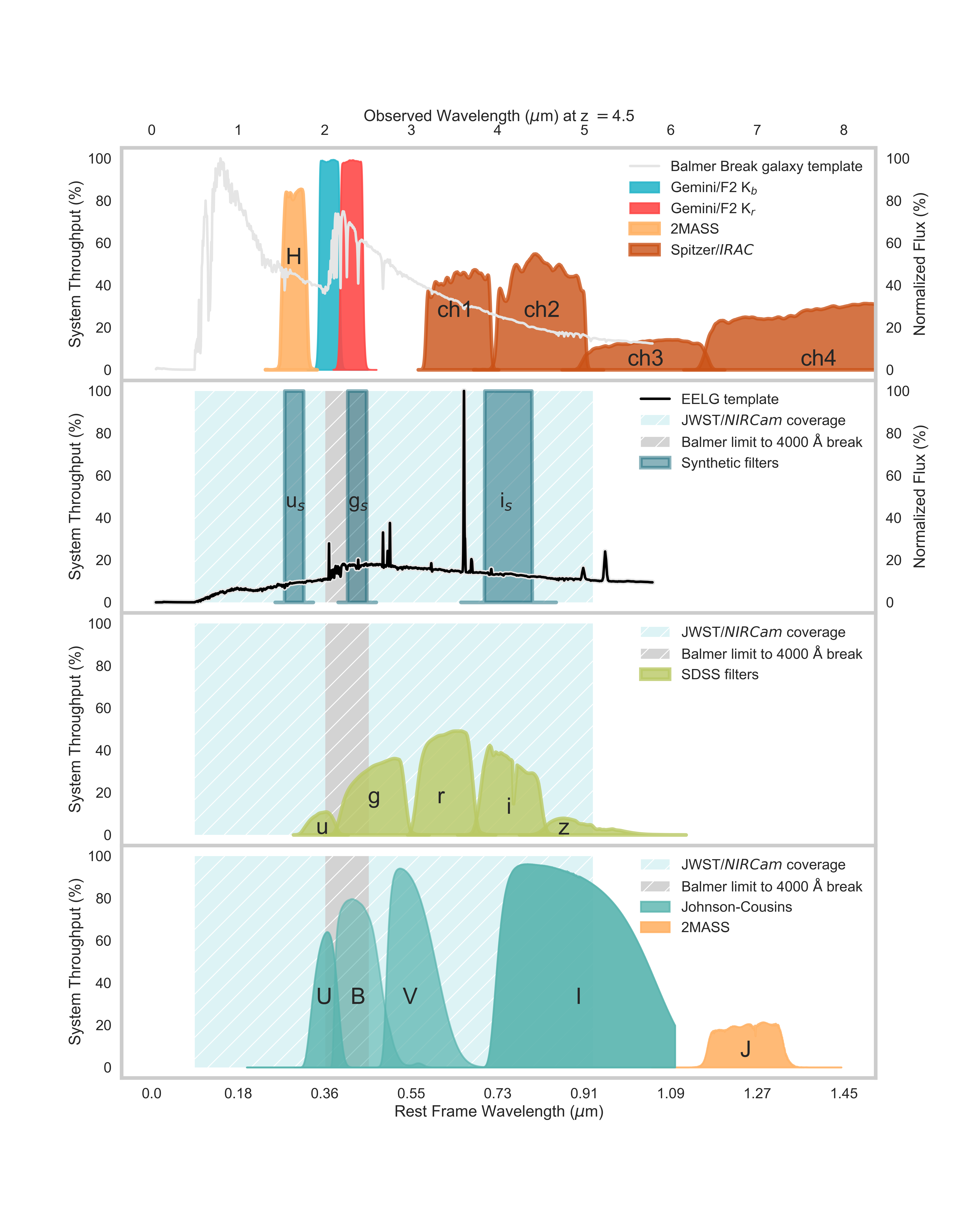}
\caption{``Classic'' photometric bands available in extragalactic catalogs and rest-frame color schemes. Rest-frame \uvj\ colors at $z >$ 3 require extrapolation because the $J$ band is redward of 5 $\mu$m (and it overlaps with \spitzer/IRAC channels 3 and 4, for which there is often limited or no data). This will be a problem even with \jwst/NIRCam observations, because they will also only extend out to 5$\mu$m. Additionally, the $V$-band band will be susceptible to artificial boosting from galaxies with strong \hb+\oiii\ emission lines (at $0.49-0.51$~$\mu$m, illustrated by the SED of the extreme-emission-line galaxy in the second panel), which comprise a significant fraction of star-forming galaxies at $z >$ 3, reaching $\sim50\%$ by z $=$ 6. We introduce a set of synthetic \sugi\ filters (second panel) designed to mitigate these problems by (1) being more sensitive to the 4000 $\AA$/Balmer break; (2) avoiding common emission lines; and (3) maintaining overlap with the \spitzer/IRAC and \jwst/NIRCam filters out to z $=$ 6.}
\label{fig:filters_all}
\end{center}
\end{figure*}

In this \textit{Paper}, we present a new set of synthetic top-hat filters to better separate galaxies in different stages (quiescent versus star-forming) using $(g_s - i_s)$ versus $(u_s - g_s)$ colors (a ``\sugi'' diagram). The outline for this \textit{Paper} is as follows.  In Section~\ref{section:data} we describe the datasets we use to test the color-selection methods.  In Section \ref{section:design} we introduce the filters ($u_s$, $g_s$, and $i_s$), which lie at at 2900 $\AA$, 4500 $\AA$, and 7500 $\AA$, respectively, and were designed to address the aforementioned issues for use with current ground-based photometric catalogs and in the \jwst\ era (Figure \ref{fig:filters_all}). In Section \ref{section:extrapolation}, we highlight the necessity for a new set of filters for use at $z >3$ by demonstrating the effects of extrapolation on \uvj. Finally, we evaluate the sample selection efficiency and limitations of each method in Section \ref{section:purity}. Throughout, we assume a $\Lambda$CDM cosmology with $\Omega_M$ = 0.3, $\Omega_\Lambda$ = 0.7 and H$_0$ = 70 km s$^{-1}$ Mpc$^{-1}$. Rest-frame colors are quoted in AB magnitudes (\citealt{oke&gunn1983}), \editone{and stellar population parameters were computed using a Chabrier IMF} (\citealt{chabrier2003}). 

\section{Data and Methods}\label{section:data}
\subsection{Observational Data ($0.5 < z < 4$)}
To test and calibrate the color-color methods, we employ four galaxy catalogs that provide both observed and rest-frame colors (we derive these ourselves where not provided), high-quality measures of photometric (or spectroscopic) redshifts, stellar masses, and star formation rates (SFR). These are important as we wish to test the ability of selection methods to identify galaxies in active and quiescent stages of their evolution.  We therefore opt to employ both theoretical catalogs constructed from known details about stellar population models (Section~\ref{subsection:simulations}) and real catalogs that contain physical high-redshift galaxies with imperfect knowledge of their stellar populations, dust attenuation, and emission line contributions (Sections~\ref{section:3dhst} and \ref{section:uvista}).  It is important to understand color-color selection efficacy from both types of catalogs as we move into the even more distant universe. 
\begin{figure*}
\begin{center}
\hspace*{-0.23cm} 
\includegraphics[trim={3cm 2cm 3.5cm 1cm},clip,scale=0.42]{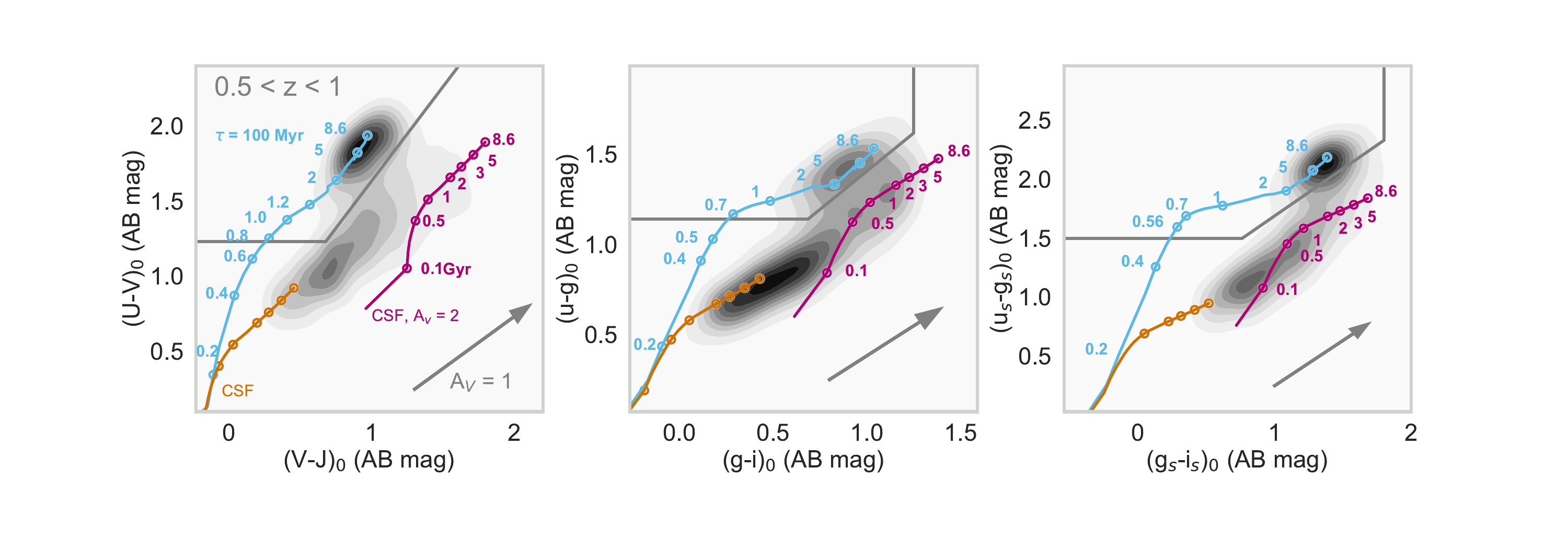}
\caption{\edittwo{Color evolution tracks of 10$^{11}$ M$_\odot$ FSPS models with various star formation histories at metallicity = 0.5 Z$_\odot$: an exponentially-declining SFH with e-folding timescale $\tau$ = 100 Myr (blue), a constant SFH with no dust (CSF, orange) and A$\rm_v$ = 2 (magenta) in \uvj\ ({\bf Left}), \ugi\ ({\bf Middle}), and \sugi\ ({\bf Right}). Empty circles denote stellar population ages in Gyr for the lower redshift boundary ($z = 0.5$). The contours show galaxies in 3D-HST at $0.5 < z < 1$. The gray vectors indicate 1 mag of extinction assuming a \cite{calzetti2000} curve. The $\tau = 100$ Myr stellar population track evolves into the quiescent region $\sim 250$ and 150 Myr younger in \sugi\ than in \uvj\ and \ugi, respectively. This suggests that \sugi\ captures recently-quenched galaxies at younger ages than the other two color selection methods.}}
\label{fig:sps_tracks}
\end{center}
\end{figure*}

\subsubsection{3D-HST}\label{section:3dhst}
One of the empirical catalogs we use are those from \threedhst\  \citep{skelton2014}, as these provide some of the most robust results from spectral-energy-distribution (SED) fitting, but are limited to $z < 2.5$ (because the catalogs are selected in the \hst/F160W band).  The data were taken in five well-studied extragalactic fields of CANDELS \citep{grogin2011,koekemoer2011} imaged in 19–45 photometric bands. We use results of \prospector\ fits to the \threedhst\ photometric catalogs from \citet{Leja2019}. The \threedhst\ catalogs provide observed-frame $0.3–24 \mu$m photometry and redshifts for $\sim 200,000$ galaxies, complete down to $10^9 M_\odot$ at $z = 2$ \citep{tal2014}.  The catalog redshifts comprise, in order of accuracy, (1) ground-based spectroscopic redshifts, (2) near-infrared grism redshifts  \citep{momcheva2016}, and (3) photometric redshifts from \eazy\ \citep{brammer2008}. 

We use stellar population parameters inferred with the $14$-parameter Bayesian SED-fitting \pa\ framework described in \citep{johnsonleja2017} and \citep{leja2017}, fit to a 90\% stellar mass-complete sample of 58461 of galaxies at $0.5 < z < 2.5$ (\citealt{Leja2019}).  The lower redshift limit ($z=0.5$) is set by the redshift at which the aperture photometry starts to become unreliable. \pa\ utilizes the wide variety of physical implementations available in \fsps\ (\citealt{conroy2010}), including a flexible six-parameter nonparametric star formation history (SFH), state-of-the-art MIST stellar isochrones, a wide range of stellar metallicities, a two-component dust attenuation model with a flexible dust attenuation curve, dust emission via energy balance, nebular line and continuum emission, and a model for the mid-infrared emission of dusty AGN tori. All of these enable \pa\ to mitigate the persistent factor-of-two uncertainty in stellar masses and SFRs derived by SED-fitting codes (\citealt{papovich2001}, \citealt{marchesini2009}, \citealt{muzzin2009}, \citealt{wuyts2009}, \citealt{behroozi2010}, \citealt{pforr2012}, \citealt{conroy2013}, \citealt{mitchell2013}, \citealt{leja2015,Leja2019}, \citealt{mobasher2015}, \citealt{santini2015}, \citealt{tomczak2016}, \citealt{carnall2018}) by making it possible to model systematics on stellar masses and SFRs galaxy by galaxy.

We note that although $UV+IR$ SFRs, which are intended to be more reliable than those from SED-fitting codes, as they account for reprocessed emission from dust-obscured star formation, are available for the \threedhst\ catalogs (e.g. \citealt{barro2019}, \citealt{whitaker2014}), they tend to underestimate the flux from old ($t > 100$ Myr) stellar populations. This becomes particularly important for low sSFR objects, with the UV and IR luminosity from old stars comprising $\sim 50\%$ of the total flux at log (sSFR $< -10.5$), and dominating the total light output at log (sSFR $= -13$) (\citealt{martis2019}, \citealt{Leja2019}). Because our study is primarily concerned with selecting highly pure samples of quiescent galaxies (which generally have low sSFRs) we chose the SFRs inferred by this modified \pa\ model, which self-consistently estimates the effect of dust heating by old stars, over catalogs based on $UV+IR$--derived SFRs.

Additionally, we compare our findings with \pa\ to those using SFRs, stellar masses, and dust extinction from \texttt{FAST} (\citealt{kriek2009}) (provided with the 3D-HST catalogs) because this has been the traditional choice in the literature. These are based on single-parameter, $\tau$-model star-formation histories, where the SFR scales as $\exp(-t/\tau)$ (for an $e$--folding timescale, $\tau$). Although FAST stellar masses are well-constrained because they mostly depend on rest-frame optical photometry, which is well covered by the 3D-HST data, star formation rates and dust are not (\citealt{wuyts2012}). We discuss this further in Section \ref{subsection:completeness_contamination}. In this work we use a subsample of galaxies with reasonable photometry (i.e. \threedhst\ {\texttt{use\_phot}} $= 1$, with $K_s$ SNR $\geq$ 7) and stellar mass M$_{*} > 10^{10}\; \rm M_\odot$.  Reported star formation rates from \pa\ are averaged over the most recent 100 Myr (rather than instantaneous values). 

\subsubsection{UltraVISTA}\label{section:uvista}

Catalogs such as CANDELS and \threedhst\ lack large samples of galaxies at the upper end of the stellar mass function, particularly at the highest redshifts (e.g., \citealt{merlin2018_1}) \editone{because they have a small survey area ($\sim 0.2$ sq degrees)}. Old and dusty galaxies tend to be better sampled by K-selected surveys. For this reason we also use results from the UltraVISTA survey (\citealt{mccracken2012}), which covers 1.5 deg$^2$ in the COSMOS field with multiwavelength data, and better samples massive galaxies at high redshifts.   For galaxies at $2.5 < z < 4$, we use MAGPHYS fits from \citet{martis2019} to the UltraVISTA DR3 catalog which is mass-complete at log(M$_*$/M$_\odot$) $> 10.5$ out to $z \simeq 4$. This deep catalog ($5\sigma$ limiting magnitude in $K_s$ = 25.2 mag) includes UV-NIR photometry spanning 49 bands. These were supplemented with  \spitzer/MIPS 24 $\mu m$ observations and 100 $\mu m$ and 160 $\mu m$ observations from the \herschel\ PACS Evolutionary Probe (PEP; \citealt{lutz2011}), and 250 $\mu m$, 350 $\mu m$, and 500 $\mu m$ observations from the \herschel\ Multi-Tiered Extragalactic Survey (HerMES; \citealt{oliver2012}). Redshifts in the catalog are either from the \eazy\ photometric redshift code or spectroscopic redshifts where available.

\citet{martis2019} used the high-z extension of MAGPHYS (\citealt{daCunha2008, daCunha2015}) to model the UV-FIR photometry of the UltraVISTA sources.  The star formation history is parameterized as a continuous delayed exponential of the form $\gamma^2$ $te^{(-\gamma t)}$, where $\gamma$ is the inverse of the star formation timescale, $\tau$. This form allows high redshift galaxies to have SFHs that rise with time (negative $t$), suggested by many observations at high redshifts (\citealt{lee2010}, \citealt{papovich2011}, \citealt{reddy2012}, \citealt{carnall2019}). Additionally, it helps reduce systematic effects introduced by too simplistic functional forms of the SFH (e.g., exponentially declining, see subsection \ref{subsection:rf_colors}). Moreover, the inclusion of the far--IR data from \herschel\ ensures stronger constraints on the estimated SFRs compared to those estimated from modeling the UV-to-NIR photometry alone. MAGPHYS SFRs are reported by averaging the most recent SFH over a 10 Myr period. For our sample, we select galaxies with $K_s$ mag $\leq$ 23.5 AB. 

\subsection{Simulation Data at $4 < z < 6$}\label{subsection:simulations}

Our goal is to develop a simple color-color selection method for distinguishing samples of quiescent and star-forming galaxies that works well for current ground- and space-based photometric data, and also in the \jwst\ era. To achieve this, we use the \jwst\ extragalactic mock catalog, publicly available as the JAdes extraGalactic Ultradeep Artificial Realizations (JAGUAR) package (\citealt{williams2018}). Unlike semi-analytic models and hydrodynamical simulations, JAGUAR uses an empirically-driven approach: modeling observed galaxy distributions and scaling relations and matching realizations to our deepest extragalactic surveys. 

The JAGUAR catalog is briefly described as follows: number counts for star-forming and quiescent galaxies are determined using the \citet{tomczak2014} mass functions and UV luminosity functions at $0 < z < 10$  (\citealt{bouwens2015}, \citealt{oesch2017}; for star-forming galaxies at $z > 4$). UV luminosity functions are used to estimate star-forming galaxy counts at $z > 4$, rather than stellar mass functions because the Balmer/4000 $\AA$ break shifts into near-infrared at these redshifts, where most facilities have low sensitivity, which could make stellar mass estimates at these redshifts uncertain. The downside to using UV luminosity functions, however, is that dusty star-forming galaxies, which are the primary contaminants of quiescent galaxies in color-selected samples, are underrepresented in UV-selected samples. We artificially infuse dusty star-forming galaxies into the JAGUAR catalog by reddening a subsample of star-forming galaxies using the \citet{calzetti2000} dust law and the mass-dependent \citet{panella2009} relation for the extinction of star-forming galaxies  in the ultraviolet (see Appendix \ref{section:appendix}). Even with this addition, we see a dearth of dusty star-forming galaxies in the JAGUAR color-color plots (discussed further in Section \ref{subsection:jaguar}). 

JAGUAR generates mock SEDs and stellar population parameters for each galaxy using \texttt{BEAGLE} (\citealt{chevallard2016}) and matches them to \threedhst\ objects based on redshift and stellar mass (for log(M$_*$/M$_\odot$) $> 8.7 + 0.4z$, which corresponds to log(M$_*$/M$_\odot$) = 9.89 at $z = 3$). Outside the parameter space covered by \threedhst, they adopt the properties of the mock SEDs from \texttt{BEAGLE}.  To derive stellar population parameters, they use a delayed exponential SFH and model stellar emission with the \citet{bc03} stellar population synthesis code. The JAGUAR catalogs incorporate line and continuum emission from gas photoionized by young, massive stars using the \citet{gutkin2016} models and dust attenuation by the two-component model of \citet{charlot&fall2000}. The catalog contains broadband photometry in \hst\ and \jwst/NIRCam filters from $0.4 - 4.8 \mu m$, emission line measurements, and galaxy morphological properties.  As with the datasets above, we use JAGUAR SFRs averaged over the most recent 100 Myr. \\ 
\vspace{-0.4em}
\subsection{Rest-Frame Colors}\label{subsection:rf_colors}
Using the aforementioned datasets (\threedhst, UltraVISTA, and JAGUAR), we derive \sugi\ rest-frame colors using \eazypy\ (\citealt{brammer2021}), a Python-based SED-fitting code based on \editone{the photometric redshift fitting code} \eazy\ \citep{brammer2008}. \editone{\eazy\ was built to handle faint galaxy samples with limited spectroscopic redshifts, as we often have with deep NIR photometric surveys. It fits a non-negative, linear combination of empirically-derived templates (in a user-defined list) to the observed photometry. Two features that distinguish \eazy\ from other photometric redshift fitting codes and make it ideal for fitting high redshift galaxies are (1) a template error function, which seeks to account for wavelength-dependent corrections of the templates, such as variations in the dust extinction law and missing spectral features; and (2) an apparent magnitude prior, which assigns low probabilities to low-redshift solutions for extremely bright galaxies at high redshift. }

\editone{One of the important modifications \eazypy\ makes to \eazy\ is that it determines rest-frame colors by doing a "weighted interpolation." It refits the templates to the data, weighting more strongly the observed photometry that is nearest the rest-frame band (in wavelength) and down-weights photometry that is farther away.  The rest-frame colors are then interpolated from the model fluxes flanking the rest-frame band of interest.}

\editone{Generally, the way the best-fit SED is determined impacts the derived rest-frame colors.   When the best fit is an arbitrary linear combination of templates (as it is in \eazy\ and \eazypy), this method produces similar results as interpolating from the observed photometry. This is important to note, as 
rest-frame colors that are based on the best-fit SED are heavily influenced by the choice of template set and the assumed star-formation history (SFH). \citet{merlin2018_1} found that rest-frame \uvj\ colors can differ by up to 0.3 mag when using an exponentially-declining  ($\tau$) SFH versus a top-hat SFH, and this change occurs mostly in the $V-J$ direction. Rest-frame colors based on the observed photometry are not subject to these systematics, however they are more sensitive to sharp effects (e.g. emission lines and spectral breaks) and low S/N. \eazypy\ adopts the best features of each method by using empirical template sets (which do not assume a SFH) and using all the available photometry, weighting more strongly bands closest to the rest-frame band of interest. This way, the estimated uncertainties on the rest-frame colors 
(e.g. Figures \ref{fig:z5galaxy}, \ref{fig:uvj_extrapolation}, and \ref{fig:ugi_extrapolation}) 
are purely statistical and not prone to the limitations and biases of the chosen template set and SFH (we return to this issue in Section \ref{section:limitations}.}

\begin{figure*}
\begin{center}
\hspace*{-0.4cm} 
\includegraphics[trim={0 0cm 0cm 0},clip,scale=0.195]{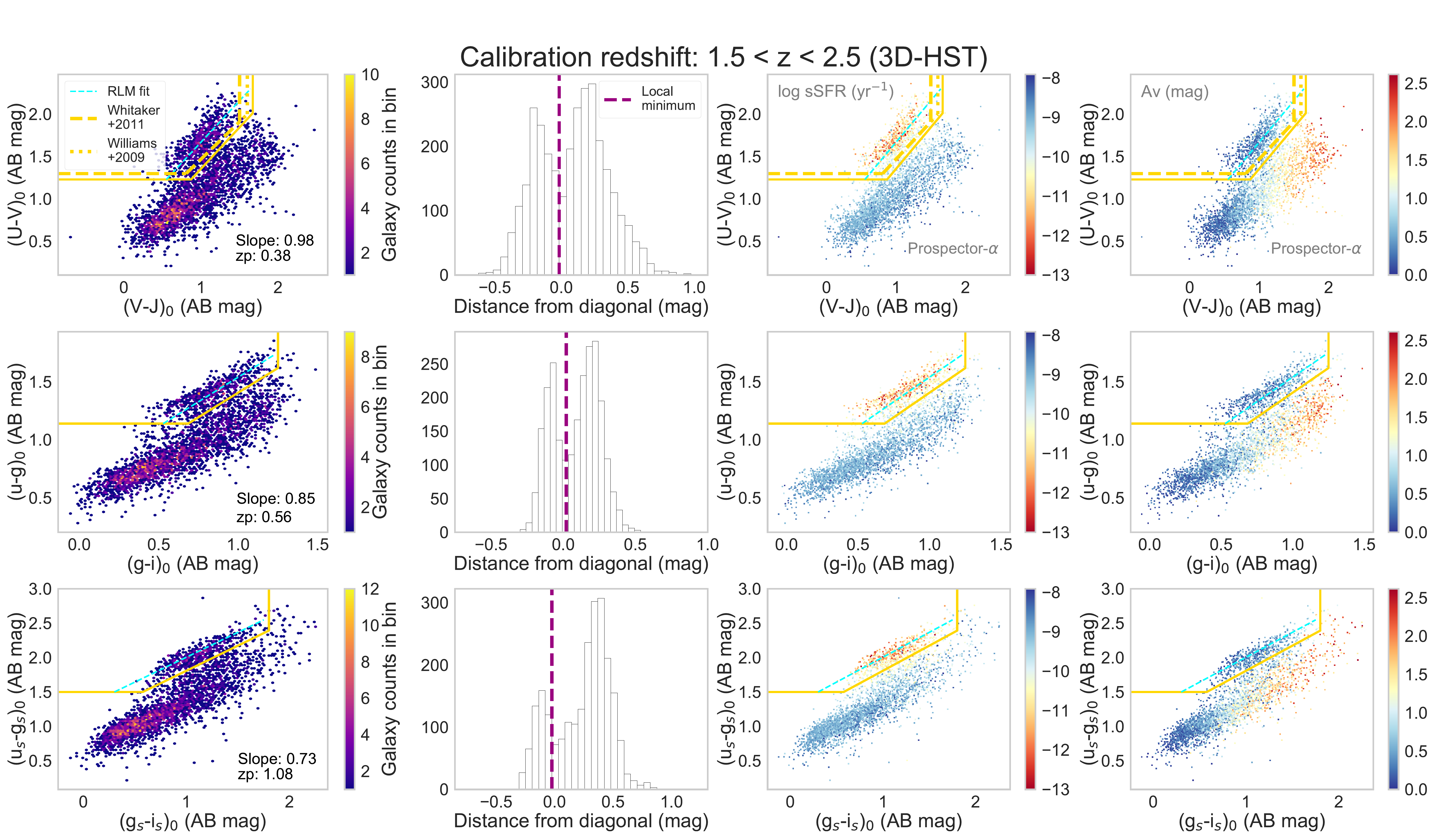}
\caption{ Calibration and validation of color-color diagrams. From top to bottom, the rows show different color-color schema  applied to samples of galaxies from the 3D-HST catalog with stellar mass, log(M$_*$/M$_\odot) \gtrsim $10, and redshift range $z = 1.5 - 2.5$. \emph{\bf Top}: \uvj; \emph{\bf Middle}: \ugi; and  \emph{\bf Bottom}: \sugi.   The \emph{\bf Left} panel of each row shows the color-color distributions.   We determine the slope of the diagonal line by fitting the ``red sequence'' (of galaxies in the upper left of each color-color space) using robust linear methods (RLM), which are less sensitive to outliers than ordinary least squares.  We then measure the y-intercept (zeropoint [zp]) by identifying a local minimum between the red and blue sequences (illustrated in the histogram in the second panel of each row). Similar to \uvj, the rest-frame \ugi\ and synthetic \sugi\ colors correlate with specific star formation rate and dust estimates from \pa, shown in the third and fourth panels in each row, respectively (dots show different galaxies color-coded by the aforementioned properties).  }
\label{fig:calibrations}
\end{center}
\end{figure*}

\editone{For catalogs that provide \uvj\ and \ugi\ colors, we use those, as they are tied to the stellar masses and star formation rates that we use for the subsequent analysis in this \emph{Paper}. This applies for instance to the \uvj\ colors from \pa\ from $0.5 < z < 2.5$ (Figure \ref{fig:calibrations}), which were determined by marginalizing over the likelihoods of the SED parameters. For catalogs that do not provide rest-frame colors (e.g., JAGUAR), we use a set of 10 templates which model the following galaxy populations: emission line galaxies, galaxies that that are both old and dusty, old quiescent galaxies, and post-starbursts. \edittwo{These templates are the same as those employed in the \threedhst\ and UltraVISTA surveys.} We used the v.1.0. template error function and fixed the redshifts to the provided photometric (or spectroscopic redshifts). } 

\edittwo{Finally, we tested the robustness of our rest-frame \sugi\ colors when derived using broadband \jwst\ photometry. This is important to consider because unlike pre-existing photometric catalogs, which have data in 25+ photometric bands (Sections \ref{section:3dhst} and \ref{section:uvista}), those from ongoing Cycle 1 programs in extragalactic fields with little-to-no ancillary data will have photometry in $\sim 10$ \jwst/NIRCAM bands. The JAGUAR catalogs provide simulated broadband \hst/ACS and \jwst/NIRCAM photometry for the JADES GTO program. This allows us to obtain realistic estimates of the uncertainties on rest-frame fluxes derived using \jwst\ catalogs at $2 < z < 6$. We find that the median fractional uncertainty ($\sigma_f/f$) for rest-frame fluxes derived using broadband \jwst/NIRCAM photometry is $<21$\% for the synthetic $i$ (\si) and SDSS $i$ bands and $<50\%$ for 2MASS $J$. Including medium-band photometry improves these uncertainties by $\sim 10\%$. }

\section{Filter Design and Calibration}\label{section:design}
\begin{table}\vspace{0.3cm}
\label{tab:table}
\caption{{Synthetic filter details. Central wavelengths ($\lambda_c$) and widths ($\Delta \lambda$) of our proposed \sugi\ filters. } \centering
\begin{tabular}{| p{0.2\textwidth}|p{0.09\textwidth}|p{0.09\textwidth}| p{0.095\textwidth}}
\arrayrulecolor{gray}\hline 
{\bf Filter Name} & $\lambda_c$ ($\AA$)& $\Delta \lambda$ ($\AA$)\\
\hline
\hline
{\bf Synthetic u (u$_s$)}  & 2900  & 400  \\
\arrayrulecolor{gray}\hline
{\bf Synthetic g (g$_s$)}  & 4500  & 400  \\
\arrayrulecolor{gray}\hline
{\bf Synthetic i (i$_s$)} & 7500 & 1000  \\
\arrayrulecolor{gray}\hline
\end{tabular}
}
\end{table}

\begin{figure*}
\begin{center}
\hspace*{-0.23cm} 
\includegraphics[trim={0cm 0cm 1cm 0cm},clip,scale=0.42]{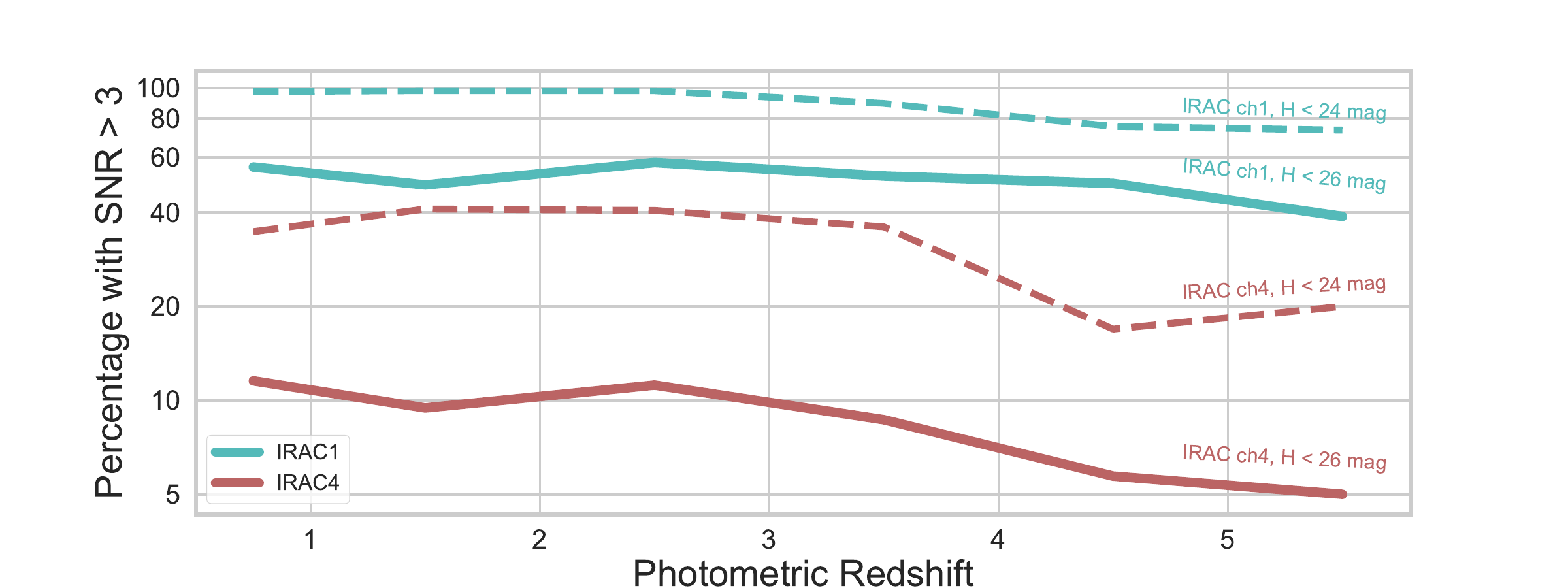}
\caption{Percentage of detected galaxies with signal-to-noise ratio greater than 3 as a function of redshift for two of the reddest bands in our deepest photometric catalogs, Spitzer/IRAC channels 1 and 4 (with central wavelengths 3.6 $\mu m$ and 7.9 $\mu m$). We use a sample of 32528 galaxies in 3D-HST at $0.5 < z < 6$ with reasonable photometry and at two different detection limits ($H$ mag $< 24$ and $H$ mag $ < 26$), with an additional requirement that the galaxy is detected in the $K$-band (SNR $>$ 5). Channels 1 and 4 correspond to the rest-frame $J$-band at z $\sim$ 2 and z $\sim$ 4.5. This means that rest frame \uvj\ colors will require extrapolation for up to $\sim$50\% of galaxies at z $\sim$ 2 and up to 95\% of galaxies at z $\sim$ 4.5. This problem will persist even with \jwst\ data, as the NIRCam filter coverage does not extend beyond 5$\mu m$ and MIRI data will be obtained for $<30\%$ of the survey area of upcoming GO, GTO, and ERS programs. The Spitzer/IRAC channels 2 and 3 data have similar detection fractions as channels 1 and 4, respectively.}
\label{fig:snr_redshift}
\end{center}
\end{figure*}

\begin{figure*}
\begin{center}
\hspace*{-0.23cm} 
\includegraphics[trim={2cm 2.5cm 3.5cm 1cm},clip,scale=0.42]{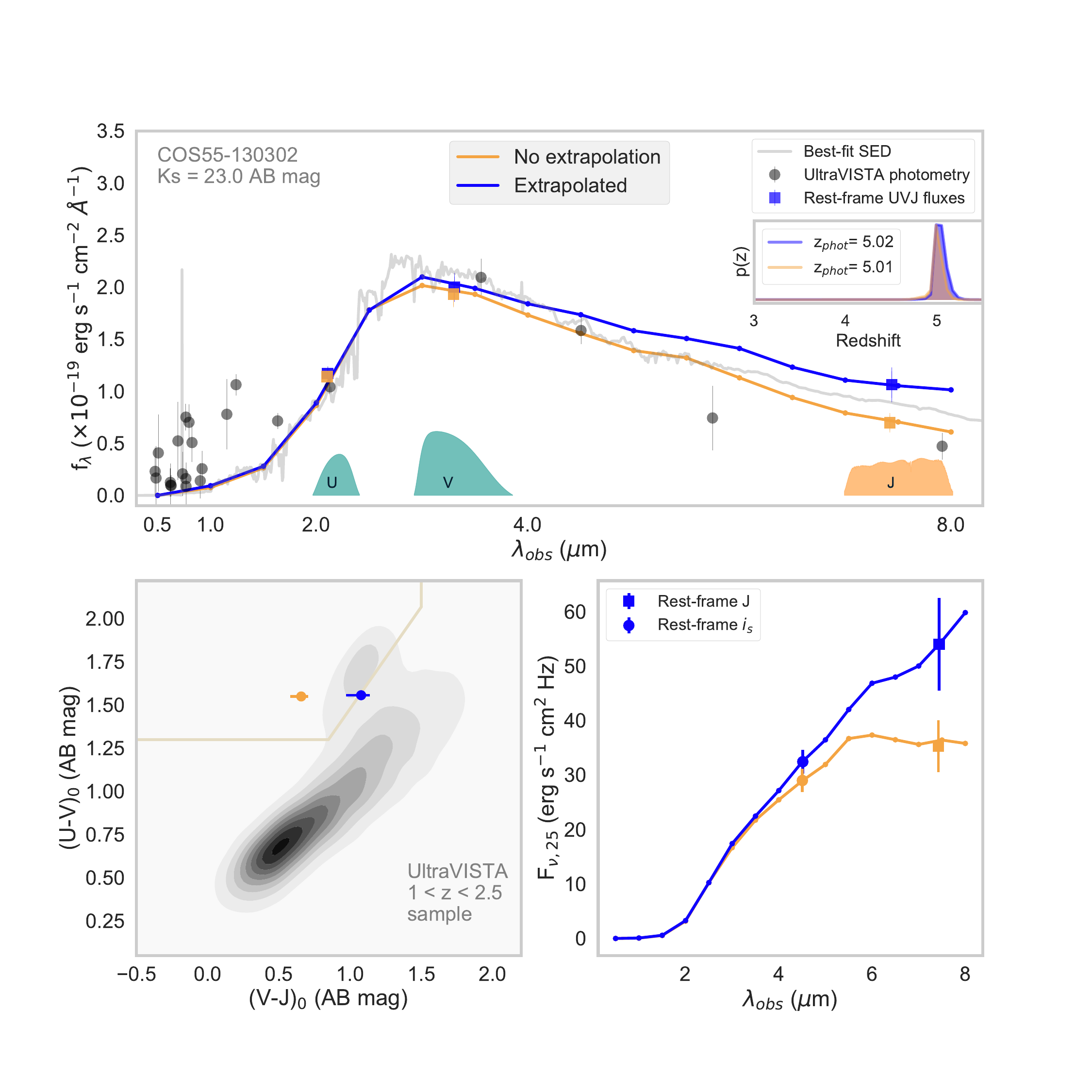}
\caption{Examining the impact of extrapolation on the rest-frame \uvj\ colors of a $z =$ 5 quiescent galaxy candidate. {\emph{\bf Top:}} Photometry, best-fit SED, and photometric redshift solution of COS55-130302 with extrapolation (blue curve/points) and without extrapolation (orange curve/points). For each case, we derive rest-frame fluxes in synthetic ($R=10$) tophat filters in the observed frame (from $0.5-8\mu m$). We simulate extrapolation by removing the observed-frame photometry corresponding to the $J$-band and redder. At  $z =$ 5, this corresponds to excluding data from \spitzer/IRAC channels 3 and 4, which is either unavailable or too shallow for at least $70\%$ of galaxies at this redshift (Figure \ref{fig:snr_redshift}). {\emph{\bf Bottom left:}} Although extrapolation has little impact on the best-fit SED and photometric redshift, it results in an almost factor of 2 difference in the estimated rest-frame $J$-band flux, resulting in a 0.46 mag change in $(V-J)_0$ color. Due to this extrapolation, galaxies like like COS55-130302 can be erroneously scattered out of \uvj-selected samples. {\emph{\bf Bottom right:}} We repeat this experiment with the \sugi\ filters and find that our ($\sg-\si)_0$ color, in comparison, is only affected by 0.08 mag, indicating that the \sugi\ diagram is better suited for selecting quiescent galaxies at $z>4$.}
\label{fig:z5galaxy}
\end{center}
\end{figure*}

\subsection{The Ideal Filter Combination}

The ideal filter set for distinguishing star-forming galaxies from quiescent galaxies at $z > 3$ should match \uvj\ in simplicity and convenience while reducing the biases inherent in \uvj\ selection such that it will be useful even in the \jwst\ era. We designed a set of top-hat filters with central wavelengths at 2900 \AA, 4500 \AA, and 7500 \AA, corresponding to the SDSS $u$, $g$, and $i$ filters, respectively, shown in Figure \ref{fig:filters_all}. 

The synthetic filters were designed to accomplish three main goals. First, they robustly straddle the Balmer/4000~\AA\ break, similar to the synthetic $U_m$ and $B_m$ filters in \citet{kriek2010}.   The wider wavelength spacing of the \su\ and \sg\ filters combined with their narrower widths means that the Balmer break produces a stronger color signature in \su$-$\sg\ than in $u-g$ and $U-V$. \edittwo{We see this in the color evolution of stellar population tracks in each diagram (Figure \ref{fig:sps_tracks}), with the $\tau = 100$ Myr model entering the quiescent region $\sim 250$ and $\sim 150$ Myr earlier in \sugi\ than in \uvj\ and \ugi, respectively.} We explore this idea further in Section \ref{subsection:beginning_end}. For this reason, the \sugi\ filters complement data from NIR medium-band surveys, which were designed to achieve higher resolution sampling of the Balmer/4000~\AA\ break.  This includes new imaging using $K_b$ and $K_r$ filters ($R\sim 10$) from the FENIKS survey, which split the $K_s$ band into a bluer and redder filter, \editone{similar to the strategy employed by the NMBS and ZFOURGE surveys, which split the $J$ and $H$ bands} for increased wavelength sampling of the Balmer break. These $K_b$ and $K_r$ filters have been shown to reduce photometric redshift uncertainties by factors of 2 and 4 at $z <3$ and $z>4$, respectively, and reduce contamination from low-redshift interlopers, usually dusty star-forming galaxies, by a factor of 2 at $4 \lesssim z \lesssim 5$ (\citealt{esdaile2021}).  We plan to apply our \sugi\ selection to data from FENIKS in a forthcoming paper. 

Second, the synthetic \sugi\ filters avoid regions with strong emission lines. These can boost photometry in a particular bandpass, mimicking red $U-V$ or blue $V-J$ colors, hence quiescence (Section \ref{subsection:eelgs}). A similar problem occurs with SDSS $u-g$ colors. We can quantify the magnitude increase ($\Delta m$) in a filter of width $\Delta \lambda$ using the following, where W$_0$ is the equivalent width of the emission line  (\citealt{papovich2001}):
\begin{equation}\label{equation:flux_boost}
\Delta m \approx -2.5 \rm log \left[1+\frac{W_0(1+z)}{\Delta \lambda}\right]
\end{equation}
A star-forming galaxy at $z = 3$, for instance, with a \editone{rest-frame} \hb+\oiii\ emission line equivalent width (EW) of 500~\AA\ will boost the $U-V$ color by $\simeq$ 1.2 mag. This problem is exacerbated at higher redshifts, because high EW objects comprise an increasingly larger fraction of the star forming population in those regimes (\citealt{endsley2021}, \citealt{tang2021}). The synthetic filters solve this problem by avoiding \oii\ (3727~\AA), \hb+\oiii\ (in the region $4863-5007$~\AA), and H$\alpha$+\nii\ (in the region $6548-6563$~\AA). However the FWHM of the \sugi\ filters is small (400~\AA\ compared to 991~\AA\ for $V$ and 1390~\AA\ for SDSS $g$). This means that although \sg\ should not be contaminated, if it is (e.g., due to photometric redshift uncertainties), a stronger magnitude change is introduced. We discuss this further in Section \ref{section:limitations}. 

Third, the synthetic \sugi\ filters improve on uncertainties from the age-dust degeneracy while avoiding extrapolation. Galaxies with blue $U-V$ colors generally exhibit unobscured star formation. Those that are red in $U-V$, however, could either be galaxies that have evolved stellar populations, or star-forming galaxies that are obscured by dust. However, because quiescent galaxies generally have relatively lower dust attenuation, they tend to have bluer $V-J$ colors. Thus, the longer the wavelength baseline, the higher the sensitivity to dust. It is for this reason that FUV/mid-infrared colors are more efficient than \uvj\ at selecting galaxies with low sSFRs (\citealt{leja&tacchella&conroy2019}).  

%
Unfortunately though, a longer wavelength baseline is more likely to require the extrapolation of observed-frame colors beyond where there is data in order to determine rest-frame colors. At $z \gtrsim$ 4, the  rest-frame $J$--band corresponds to $\gtrsim$ 6~\micron\ observed, generally only available from IRAC 5.8 $\mu m$ and redder bands (e.g. forthcoming \jwst/MIRI data). At longer wavelengths, the flux sensitivity of the IRAC data is substantially lower (by factors of 4 and 8 in IRAC channels 3 and 4 compared to IRAC channels 1 and 2), hence those data are \editone{often unavailable or too shallow} (Figure \ref{fig:snr_redshift})  \footnote{IRAC Instrument Handbook: \url{https://doi.org/10.26131/irsa486}}. Rest-frame \uvj\ colors at $z> 4$ therefore require extrapolation and hence are less reliable. Figures \ref{fig:z5galaxy} and \ref{fig:uvj_extrapolation} illustrate the effects of extrapolation on the $J$-band. The new synthetic filters proposed here cover the longest wavelength baseline that is possible while maintaining overlap with \spitzer/IRAC channels 1 and 2 and \jwst/NIRCam coverage, to avoid extrapolation at redshifts out to $z\sim 6$ (the end of the epoch of reionization).  We demonstrate the effects of extrapolation on colors using the rest-frame $J$-band further in Section \ref{section:extrapolation}. 

Finally, we discuss the choice of a synthetic \su\ filter over an ultraviolet one (e.g., rest-frame $NUV$). $NUV-r$ or $FUV-V$ color selection, for instance, is a better discriminator of current versus past star formation activity than $U-V$ (\citealt{ilbert2010}, \citealt{martin2007}, \citealt{arnouts2007}    , \citealt{ilbert2013},\citealt{man2016}, \citealt{hwang2021}, \citealt{leja&tacchella&conroy2019}). Similarly, the $FUV-V$ color has also been shown to have a much stronger correlation with specific star formation rate (\citealt{leja&tacchella&conroy2019}). NUVrJ-selected quiescent samples have also been shown to have low contamination ($\leq$ 10-15\%) from dusty star-forming galaxies at $z\leq$ 3 (\citealt{man2016}, \citealt{hwang2021}). These are all reasons to choose an ultraviolet filter over an optical $u$ filter. 

However, a rest-frame color selection involving a UV filter 
would miss recently-quenched or "post-starburst" galaxies, particularly at high redshifts. \editone{Post-starbursts have intermediate-age stellar populations, which means that while they have strong Balmer/4000 \AA\ breaks, they also have a higher UV continuum than that of a typical quenched galaxy due to residual star formation (see the template in the first panel of Figure \ref{fig:filters_all}). As a result of their elevated UV flux, their $NUV-r$ colors are \emph {blue}, similar to those of star-forming galaxies. This has been demonstrated in the literature.} For example, \citet{valentino2020} showed two post-starburst galaxies at $z\sim 4$ that would be classified as star-forming using $NUVrJ$-selection and quiescent using \uvj. 

By extension, post-starburst galaxies may be missed using rest-frame $NUV-g$ colors.  These galaxies are observed (and expected) to be more common at higher redshifts ($z > 3$ e.g. \citealt{forrest2020}, \citealt{d'eugenio2020}). It is therefore prudent to use a color-selection that will include them. For this reason, the rest-frame $\su - \sg$ color is better suited to identify galaxies with colors typical of post--starburst galaxies, which will be an important class of objects at $z > 3$.   

\subsection{ Color-Color Selection Criteria}
In this section, we describe how we define empirical \uvj, \ugi, and \sugi\ selection criteria that separate quiescent and star-forming galaxies in our data. \editone{The basic quantities that we are interested in are the slope and y-intercept of the diagonal line, as well as the positions of the horizontal and vertical color cuts. These are based on a fit to an initial quiescent selection, which we vary until we obtain a clear bimodal distribution centered at zero. This process is a modification} of the method in \citet{kawinwanichakij2016}. It is summarized as follows, and illustrated in Figure~\ref{fig:calibrations}:
\begin{enumerate}
\itemsep -0.3em 
    \item Find quiescent galaxies using literature criteria for the specified redshift where available (e.g. \citet{williams2009}, \citet{whitaker2011}, \citet{muzzin2013} for \uvj) or by making an initial guess based on the distribution of data in color-color space (for \ugi\ and \sugi). 
    \item Update the slope of the diagonal line using the slope from an iteratively re-weighted robust linear methods (RLM) fit to a sigma-clipped subsample of the galaxies in the ``red sequence'' region of the color-color space (these are candidate  quiescent galaxies). We opt to use RLM fits as they are less sensitive to outliers and do not assume homoscedastic errors (e.g. \citealt{rlm}).  This is illustrated in the first panel of each row of Figure~\ref{fig:calibrations}.   
    \item Find the distance of each point in the data set from the diagonal line to delineate the red and blue sequences.  This is illustrated in the second panel of each row of Figure~\ref{fig:calibrations}.   
    \item Bin distances to determine the local minimum between the two sequences. The optimal number of bins is determined using the Freedman-Diaconis rule (\citealt{freedman&diaconis}), which minimizes the area under the probability distribution function of the data and that of the function that best matches the data. Finally, we use the local minimum to update the y-intercept of the diagonal line.
\end{enumerate} 

We evaluate our method first by applying it to a sample of galaxies from the ZFOURGE survey \citep{straatman2016} and the third data release of the UltraVISTA survey at $0.3 < z < 1.6$, following \citet{kawinwanichakij2016}. Our zeropoints and slopes for the aforementioned are within 0.02 and 0.17~mag of \citet{kawinwanichakij2016} for the two datasets, respectively. 

\editone{We then determine the color-color selection criteria for each dataset used in this paper (\threedhst, UltraVISTA, and JAGUAR) using the method outlined above. This is important, as the color lines are not portable from one survey to another due to systematic biases (discussed further in Section \ref{subsection:completeness_contamination}).} Because the observed data at $z > 3$ are sparse and generally of poorer quality, we calibrate the color-color lines on a broad redshift range ($z = 1.5 - 2.5$) and apply these calibrations to the entire dataset, up to $z = 6$ in the discussion that follows. \editone{Calibrating the lines on a broad redshift range enables us to account for any evolution in the zeropoint as a function of redshift. We apply a zeropoint adjustment at $z = 4-6$ based on the \uvj\ colors of spectroscopic samples at z = $3-4$ in the literature (Section \ref{subsection:z4results})}.

Figure~\ref{fig:calibrations} shows a bimodality in each panel, indicating the presence of a distinct star-forming and quiescent population in all three color-selection methods: \uvj, \ugi, and \sugi. We define the selection of quiescent galaxies for each of the color-color selections in Figure~\ref{fig:calibrations} with the following criteria.  

\smallskip
\noindent For \uvj:  
\begin{equation}
  \begin{split}
  (U-V)  > 1.23  {~~~~~}{~~~~~} \land {~~~~~}{~~~~~} (V-J)  < 1.67 & {~~~~~}{~~~~~}\land \\  
     (U-V)~~~~ >~~~~  (V-J) \times 0.98 + 0.38  ~~~~~~~ & 
   \end{split}
\end{equation}
For \ugi:
\begin{equation}
  \begin{split}
    (u-g) > 1.14 {~~~~~}{~~~~~} \land 
    {~~~~~}{~~~~~} (g-i) < 1.25 & {~~~~~}\land \\  
    (u-g)~~~~ >~~~~ (g-i) \times 0.85 + 0.56 ~~~~~~~ &
    \end{split}
\end{equation}
For \sugi:
\begin{equation}
  \begin{split}
    (u_s - g_s) >&1.5 {~~~~~}{~~~~~}  \land   {~~~~~}{~~~~~} (g_s - i_s) < 1.8  
    {~~~~~}\land \\
    (u_s - g_s)~~~~ >~~~~&(g_s - i_s) \times 0.73 + 1.08 ~~~~~~~ 
    \end{split}
\end{equation}
In all cases, $\land$ is the logical AND operator. \\

\section{Effects of Extrapolation}\label{section:extrapolation}
\begin{figure*}
\begin{center}
\vspace*{-2.4cm} 
\hspace*{-0.8cm} 
\includegraphics[trim={0 7cm 0 0},clip, scale=0.33]{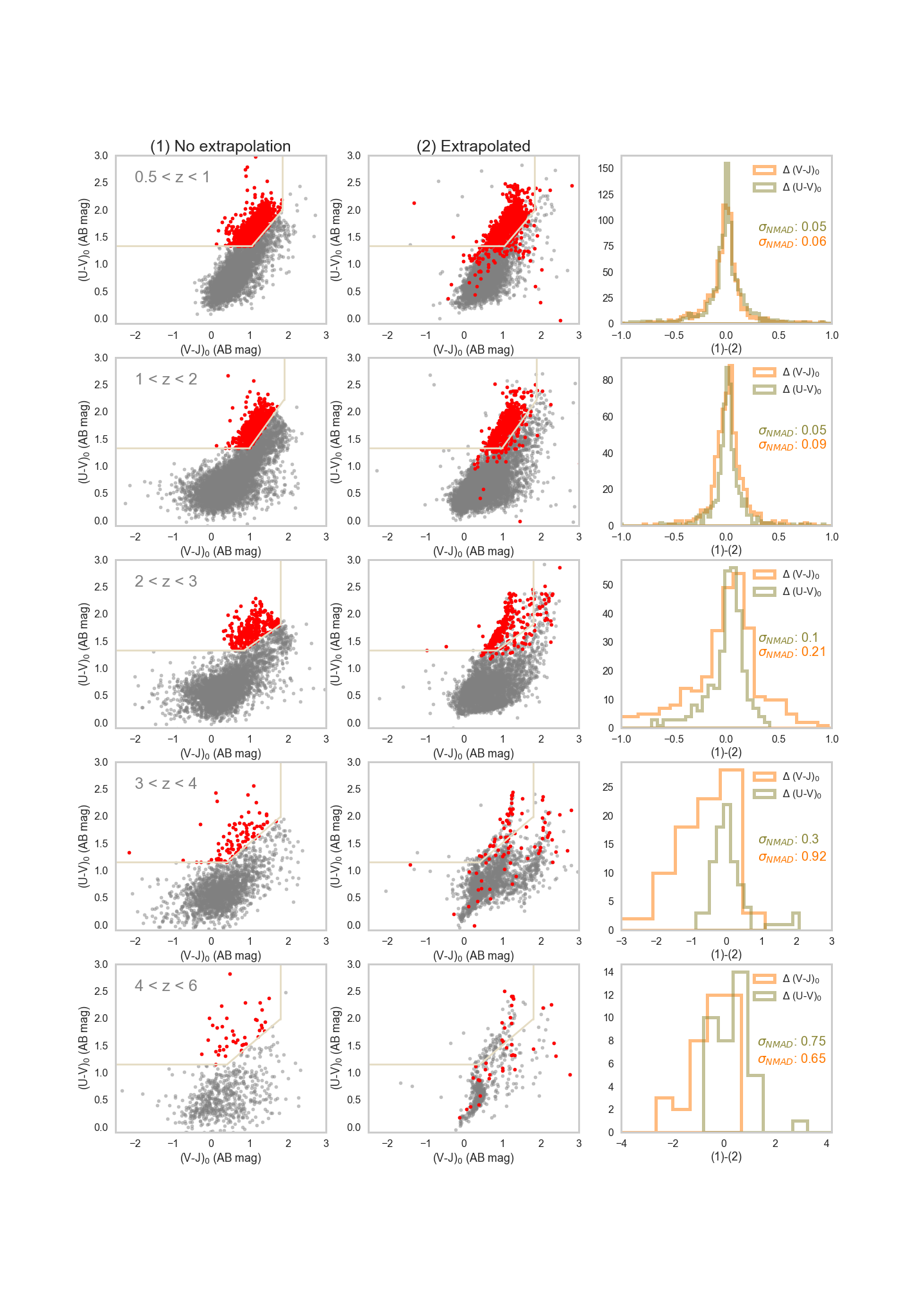}
\caption{Impact of extrapolation on \uvj-selected quiescent galaxies \editone{with rest-frame $J$-band coverage from \spitzer} at $0.5 < z < 6$.  Each row shows the galaxies in bins of increasing redshift (top to bottom).   The {\emph{\bf Left}} panel in each row shows the distribution using rest-frame colors derived with all the available photometry (no extrapolation). The red dots show galaxies selected as $UVJ$-quiescent. The {\emph {\bf Middle}} panel of each row shows the colors of the same galaxies on the left when the rest-frame colors are extrapolated (when the observed fluxes corresponding to the $J$-band and redder are removed).   The {\emph{\bf Right}} panel of each row shows the distributions of the difference in $U-V$ and $V-J$ colors between the non-extrapolated and extrapolated cases for the quiescent galaxies. We quantify this change using the  normalized median absolute deviation ($\sigma_{NMAD}$). The effects of extrapolation are seen as an increase in $\sigma_{NMAD}$ with increasing redshift, with $\sigma(\Delta[V-J])$ increasing from 0.2~mag at $z\sim 2$ to $>$0.65~mag at $z > 3$.   NB: the abscissa ranges in the top three right-hand plots are the same, however we adjust it for the bottom two ($z > 3$) to better show the large change extrapolation introduces to rest-frame colors at these redshifts.  }
\label{fig:uvj_extrapolation}
\end{center}
\end{figure*}

\begin{figure*}
\begin{center}
\vspace*{-2.4cm} 
\hspace*{-0.8cm} 
\includegraphics[trim={0 7cm 0 0},clip, scale=0.34]{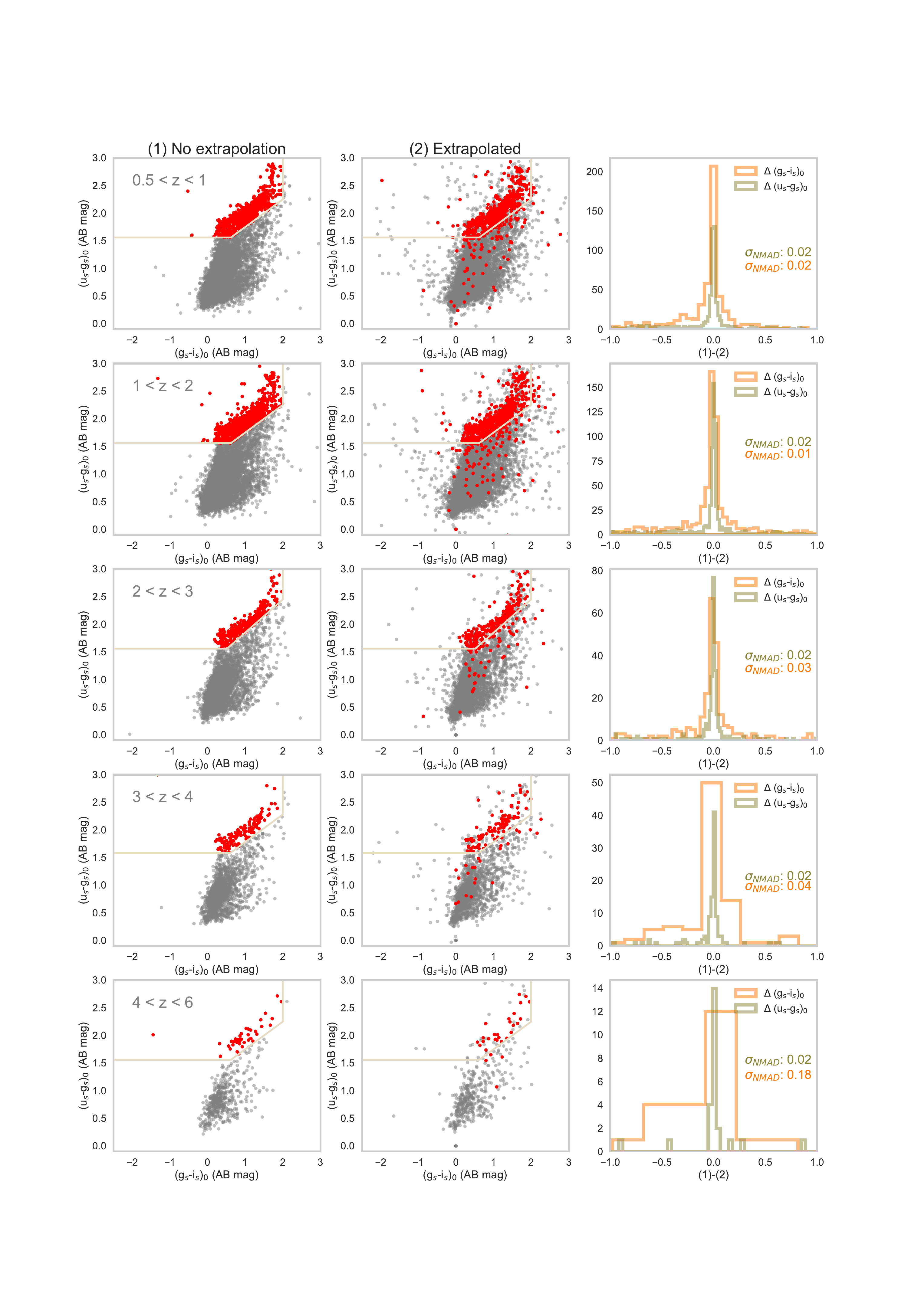}
\caption{Similar to Figure~\ref{fig:uvj_extrapolation}, but showing the impact of extrapolation on synthetic \sugi-selected quiescent galaxies at $0.5 < z < 6$. The \sugi\ filters are more resistant than \uvj\ to the effects of extrapolation, particularly at $z > 4$.  The scatter in the distribution of the change in $(g_s - i_s)_0$ (\emph{\bf Right} panels in each row) increases only slightly with redshift and is never higher than $\sigma(\Delta [g_s - i_s])$ = 0.2 mag.  This is significantly improved compared to the effects of extrapolation on the $UVJ$ colors.  NB: the limits of the abscissa of the right-most panels are the same in all rows and the same as that of the first three rows of Fig~\ref{fig:uvj_extrapolation}.}
\label{fig:ugi_extrapolation}
\end{center}
\end{figure*}

\uvj\ colors at z $>$ 2.5 can be uncertain because the rest-frame $J$-band is subject to extrapolation. \editone{Generally}, rest-frame colors are determined by interpolating between the two filters flanking the rest-frame band of interest (see Section \ref{subsection:rf_colors} for a more detailed description). At $z=3$, for instance, the $J$-band overlaps with \spitzer/IRAC channels 1 and 2 at $3.6 \mu m$ and $4.5 \mu m$, respectively. These data are often available (at SNR $>$ 3) for only $\sim 55\%$ of galaxies in our deepest photometric catalogs (Figure \ref{fig:snr_redshift}). This means that for $\sim 45$\% of galaxies at $z \sim$ 3, their rest-frame $J$ fluxes are determined by extrapolating from the reddest observed data point. This problem is exacerbated at higher redshifts ($z \gtrsim 4$) becaus longer wavelength data (channels 3 and 4) are available for an even smaller fraction of galaxies ($< 7\%$). 

The availability of mid-IR data with the {\emph {James Webb Space Telescope}} will not solve this problem. At the beginning of its mission, \jwst\ will carry out a number of General Observer (GO),  Guaranteed Time Observations  (GTO), and Early Release Science (ERS) surveys that are aimed at studying galaxies at high redshift. The largest programs (in order of decreasing survey area) are COSMOS-Web (\citealt{kartaltepe2021}), Public Release IMaging for Extragalactic Research (PRIMER; \citealt{dunlop2021}), the \jwst\ Advanced Deep Extragalactic Survey (JADES; \citealt{williams2018}), and the Cosmic Evolution Early Release Science survey (CEERS; \citealt{finkelstein2017}). Together, these programs cover the five standard extragalactic fields (COSMOS, UDS, GOODS-N, GOODS-S, and EGS). Although they all include MIRI imaging, mid-IR data will be available for only a small portion of the surveyed area, due to the instrument's relatively small field of view. COSMOS-Web and PRIMER, which together survey COSMOS and UDS, will have MIRI coverage for up to $\simeq$ 30\% and $\simeq$ 60\% of their planned survey areas, respectively (noting that the survey area of PRIMER is only 1/6 that of COSMOS-Web). \editone{Additionally, the majority of planned MIRI observations will be shallow, i.e. 5$\sigma$ depth $\lesssim$ 24 AB mag. For reference, the candidate $z = 5$ quiescent galaxy in Figure \ref{fig:z5galaxy} has m$_{IRAC 4}\approx$ 25 AB mag.} The latter two surveys, which cover GOODS-N, GOODS-S, and EGS, will have MIRI coverage for $7.4\%$ and 25\% of their total area, respectively. This means that we will have adequate MIRI coverage for only two out of the five extragalactic fields, and consequently a small fraction of galaxies surveyed. As such, extrapolation will continue to be a problem for the majority of galaxies at $z > 3$, even with \jwst\ data. 

We illustrate this with a candidate quiescent galaxy in the UltraVISTA catalog with a photometric redshift at $z_\mathrm{ph} \simeq 5$ (\jwst\ GO Program \#2362, \citealt{marsan2021}).  This galaxy offers an excellent example because it includes many high S/N photometric points from the UV-to mid-IR, such that the true SED is well-represented by the data. We simulate extrapolation by removing photometric points corresponding to the rest-frame $J$-band (and redder) at that redshift.  We then re-derive the photometric redshift and rest-frame colors from the data using \eazypy\ (using the same method described in Section \ref{subsection:rf_colors}).  We determine the rest-frame \uvj\ fluxes and corresponding uncertainties in the ``non-extrapolated'' case (where we use all available photometric bands) and the ``extrapolated'' case (where we have excluded bands corresponding to the rest-frame $J$-band or redder, specifically IRAC ch3 and 4 in this case).  This latter case mimics the effects of extrapolation on the derived rest-frame \uvj\ fluxes, shown in Figure \ref{fig:z5galaxy}. We also show the rest-frame colors derived in synthetic top-hat ($R=10$) filters spanning $0.5 - 8 \mu m$ for both cases. 

Extrapolation (removing data points covering the rest-frame $J$-band and redward) has a large impact on the derived rest-frame $J$-band flux (though it does not significantly change the redshift, which is constrained by the data near the Balmer/4000~\AA\ break).  In this case, the rest-frame $J$-band flux is changed by a factor of 2, causing the $V-J$ color to become redder by $\simeq$0.5 mag.  This moves the galaxy out of the quiescent region of the $UVJ$ diagram. We repeat this experiment with the \sugi\ filters (removing observed data corresponding to the \si-band and redder) and find that the ($\sg-\si)_0$ color, in comparison, is only affected by 0.08 mag, indicating that the \sugi\ diagram is more robust against extrapolation and hence better suited to select quiescent galaxies at $z>4$ with incomplete data. 

Our findings with COS-130302 beg the question: how often do studies misclassify galaxies as quiescent using the \uvj\ diagram due to extrapolation? We investigate this by repeating the experiment above on a SNR-limited sample ($K_s$ SNR $\geq$ 5) of galaxies at $0 < z < 6$ from the 3D-HST survey. This is illustrated in Figure~\ref{fig:uvj_extrapolation}.  We quantify performance by measuring the normalized median absolute deviation ($\sigma_{NMAD}$, \citealt{beers1990}, \citealt{brammer2008}) of the change in rest-frame colors ($\Delta (U-V)_0$ and $\Delta(V-J_0)$) of quiescent galaxies in the ``non-extrapolated'' (using all data) and ``extrapolated'' cases (removing any observed data corresponding to and redward of the rest-frame $J$-band). 

\uvj\ selection performs very well out to $z\sim$ 2. For $z \lesssim 2$ the difference between non-extrapolated and extrapolated colors of quiescent galaxies is small, with $\sigma[\Delta (V-J)_0] \lesssim 0.1$~mag. At $z >$ 2, the scatter increases to $\sim$ 0.3 mag, and it becomes much worse at $z>4$, reaching $\sigma[\Delta (V-J_0)] \sim 1$~mag at $4 < z < 6$. $(U-V)_0$ colors also start to become extrapolated at $z \sim 2$, where the rest-frame $V$-band enters the observed $H$-band, which is only detected at SNR $>$ 3 for up to $\sim$45\% of galaxies at $z \gtrsim$ 2 (estimated for Figure \ref{fig:snr_redshift}, not shown). This likely explains the factor of 2 increase in $\sigma_{NMAD}$ (0.05 $-$ 0.1) from $1<z<2$ to $2< z<3$ in $(U-V)_0$. 

In contrast, the effects of extrapolation on the synthetic rest-frame \sugi\ colors are much lower, even at high redshift.  Figure~\ref{fig:ugi_extrapolation} shows the results.  The change in $\Delta\sugi$ colors remains small between the extrapolated and fiducial cases, with $\sigma[\Delta\sugi] \leq 0.2$~mag at all redshifts, even at $4 < z < 6$.\footnote{The effects of extrapolation on $U-V$ color may be reduced with \jwst/NIRCam data because unlike with ground-based telescopes, a number of narrow, medium, and broad-band filters exist between the $K$-band and IRAC channel 1. Nevertheless, these effects will likely persist for $V-J$ colors.} Therefore, we conclude that extrapolation of the rest-frame $J$-band can have a serious impact on the selection of quiescent galaxy candidates at high redshifts ($z > 4$).  This is especially important as the quiescent fraction is expected to decline sharply at this epoch (e.g., \citealt{merlin2019}) such that these types of systematics of selection may dominate searches for such objects (see discussion in Section \ref{section:purity} below). 

We note that these results are an illustration of the effects of extrapolation on the selection of high redshift quiescent galaxies in the specific case where rest-frame colors are derived \editone{semi-empirically}. When they are estimated \editone{solely} from the best-fit SED, as is typically done in the literature, the effects of extrapolation on a single galaxy are varied and dependent on many assumptions, including its intrinsic SED shape, SNR, redshift, the assumed star formation history, and the template sets and parameters used for SED-fitting. Our experiments demonstrate the need for an updated color-selection method for high redshift galaxies.  This will continue to be necessary in the \jwst\ era. We have demonstrated that the effects of extrapolation on our proposed synthetic \sugi\ filters will be minimal out to $z\sim$6. 

In the remaining sections, we show that the synthetic \sugi\ filters match \uvj\ in simplicity and discrimination efficiency, making \sugi\ a viable replacement.  

\section{Purity Tests and Discussion}\label{section:purity}
\begin{figure*}[h]
\begin{center}
\vspace*{-1.5cm}
\includegraphics[trim={0 8cm 0 0},clip,scale=0.3]{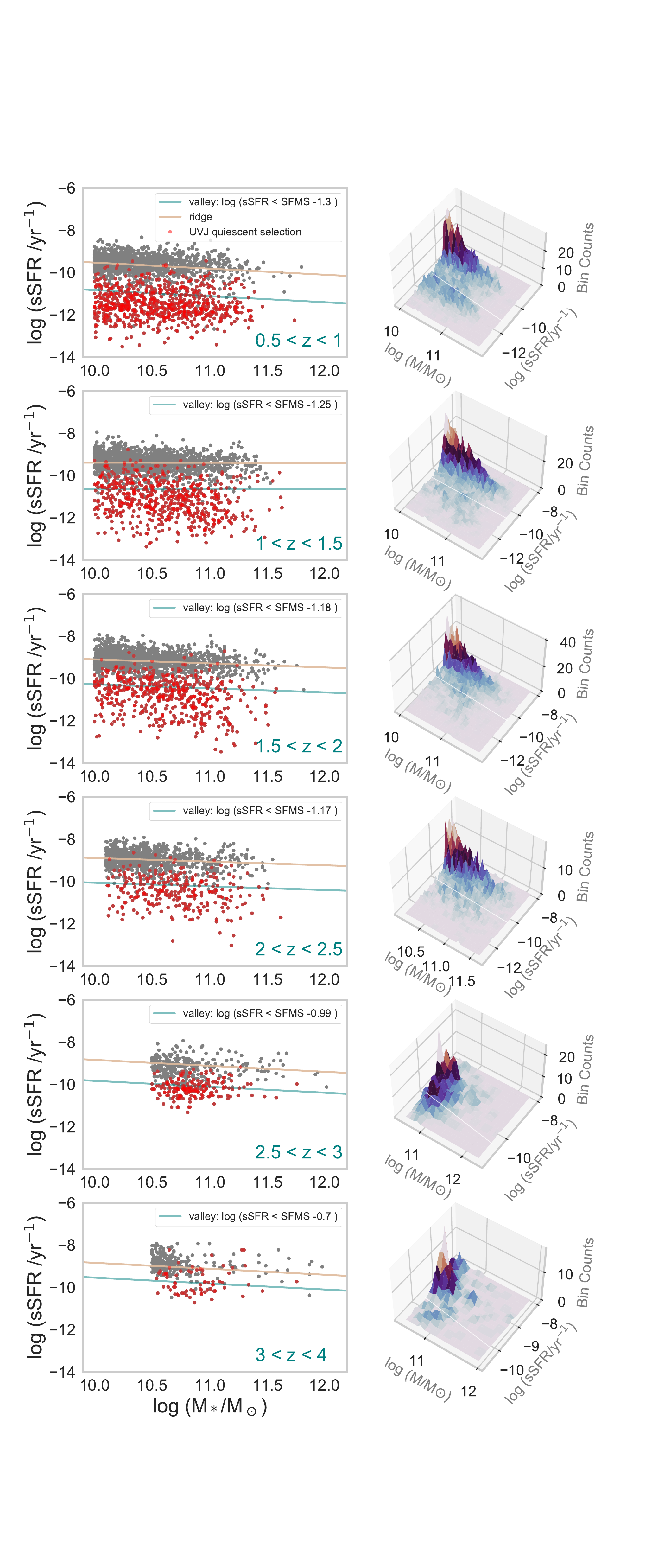}
\caption{Selection of quiescent galaxies based on specific star formation rate (sSFR). Each row shows the distribution of sSFR as a function of stellar mass in bins of redshift (low--to--high, top--to--botoom).  In each row,the \emph{\bf Left} panel shows the selection of quiescent galaxies by identifying the density peak of the star forming main sequence (SFMS) at each redshift and the green valley between the red and blue sequences, which we define as where the data falls below 10\% of the peak. Because our galaxy samples at $z > 2.5$ are small, we use the slope of the SFMS at $2 < z < 2.5$ for the two highest redshift bins. The mass range at each redshift reflects the 90\% stellar mass completeness limit of the 3D-HST ($0.5 < z < 2.5$) and UltraVISTA ($2.5 < z < 4$) surveys.  The \emph{\bf Right} panel in each row then shows a 3D projection of the SFMS at each redshift. The white line shows the estimated location of the green valley between the red and blue sequences. }
\label{fig:sfms}
\end{center}
\end{figure*}

\begin{figure*}
\begin{center}
\includegraphics[trim={0cm 0cm 0cm 0cm},clip,scale=0.36]{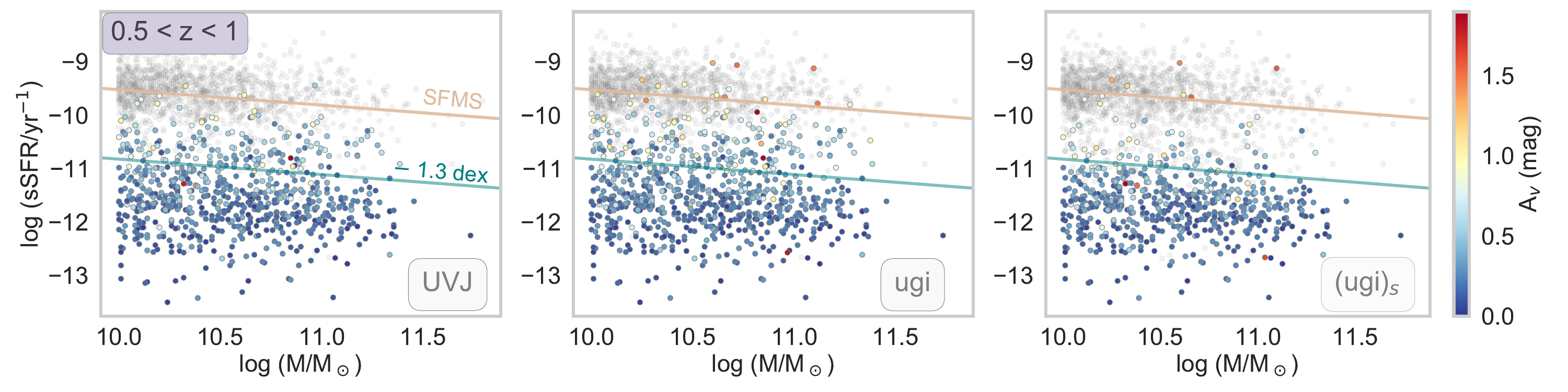}
\caption{Demographics of quiescent galaxies selected in each method at $0.5 < z < 1$ in \threedhst. The colored points are the galaxies identified as quiescent in \uvj, \ugi, and \sugi. We color code each sample by dust extinction from \pa. Gray points mark galaxies outside of the quiescent region. The "green valley" (teal line) and peak of the star forming main sequence (brown line) are the same as those from Figure \ref{fig:sfms} at this redshift. Due to a slightly smaller baseline than \uvj, \ugi\ and \sugi\ select a few galaxies with high dust obscuration (A$_{\rm V} \geq 1$), as these tend to lie closer to the diagonal line in these color diagrams than in \uvj. The majority of false positives (galaxies above the teal line) have moderate dust extinction values (median A$_{\rm V}$ $\approx$ 0.5) and high specific star formation rates ($< -10 \rm\; yr^{-1}$). \sugi-selected quiescent samples have less contamination from these dusty galaxies than \uvj\ and \ugi.}
\label{fig:demographics}
\end{center}
\end{figure*}

\subsection{Tests of Completeness and Contamination using Color-Color Selection}\label{subsection:completeness_contamination}
Color-color selections involve a balance between sample purity and completeness. Generally, the higher the completeness, the higher the number of contaminants and vice versa. This tradeoff becomes increasingly important for quenched galaxies at higher redshift ($z > 3$),  where features such as the Balmer/4000 $\AA$ break, which are key for determining photometric redshifts, are shifted into the near-infrared, a regime prone to numerous systematics. Additionally, galaxies at those redshifts generally have lower signal-to-noise photometry, which can decrease the accuracy of rest-frame colors. As a result, dusty star forming galaxies \footnote{\editone{These can either be dusty galaxies \emph{at the same redshift} as the quiescent galaxies or at lower redshift whose colors mimic those  of high-redshift quiescent galaxies. Most of the dusty contaminants, however, will fall into the former category. The placement of the \sugi\ filters along the Balmer/4000 \AA\ break will help reduce the effect of the latter (see Section \ref{section:design}).}}, which are the primary contaminants of quiescent samples, are more likely to scatter into the quiescent region.  Furthermore, galaxies with strong nebular line emission (e.g.,  strong \oiii\ emission) are another source of contamination, and such sources comprise a greater fraction of the star-forming population at these redshifts \citep{endsley2021}. Efficient color-color selection of galaxies at these redshifts therefore requires careful culling of such objects in order to maintain high completeness while lowering the contamination rate. In this section, we contrast the performance of the \sugi\ diagram with that of \ugi\ and \uvj\ as it pertains to identifying and removing such interlopers. 

One major challenge is evaluating ``truth'' in the purity and contamination of quiescent-galaxy searches. Several approaches have been taken in the literature, including using estimates of galaxy star formation rates (and specific SFRs) (e.g. \citealt{heinis2014}, \citealt{chang2015}, \citealt{tasca2015}, \citealt{tomczak2016}, \citealt{pacifici2016}). Others identify active galaxies based on emission line strength (e.g. \citealt{rodighiero2014}, \citealt{belfiore2018}) or far-IR emission (e.g. \citealt{whitaker2014}, \citealt{lee2015}, \citealt{ilbert2015}, \citealt{schreiber2015}, \citealt{erfanianfar2016}).  \citet{speagle2014} and \citet{popesso2019} showed that the slope of the SFR-M* plane, commonly referred to as the star-forming main sequence (SFMS), changes by up to 1 dex depending on which of these methods is used. Similar results were found with the Illustris-TNG simulations, with quiescent fractions on the higher mass end (log(M$_*$/M$_\odot$) $> 10.5$) varying between $20-40\%$ (\citealt{donnari2019}). Moreover, these thresholds only correspond to a “relative” quiescence because they fail to account for the evolution in the intrinsic stellar populations of quiescent galaxies over cosmic time, making them applicable only to a specific redshift or stellar mass bracket. Additionally, while $UV+IR$ SFRs, which some regard as the "gold standard", are available for our dataset at $z < 4$,  these can be problematic for a number of reasons. This is because IR luminosities (typically estimated from the 24 $\mu m$ flux) tend to overestimate SFRs by up to a factor of 4, because they do not account for heating from intermediate age and old stars, which tends to dominate the IR budget for log(sSFR) $< - 13$ (\citealt{martis2019}, \citealt{Leja2019}, \citealt{belli2017}, \citealt{fang2018}). Simulations also report similar findings, i.e. that UV luminosities can be overestimated by $0.2-0.5$ dex at $z = 1-4$ for galaxies with log(M$_*$/M$_\odot$) $\approx 8.5 - 10.5$ (\citealt{katsianis2020}) and that the total IR luminosity is not a good proxy for dust-obscured star formation for rapidly-quenched or "post-starburst" galaxies (\citealt{roebuck2019}).  In summary, the definition of what constitutes ``truth'' in quiescent-galaxy selection matters. 

To mitigate this effect, we choose a data-driven approach. In order to determine which galaxies are truly quiescent, we take SFR and stellar mass estimates for galaxies from two catalogs \editone{that account for these systematics by self-consistently modeling the entire SED}:  the \citet{Leja2019} measurements from \pa\ for 3D-HST galaxies at $0.5 < z < 2.5$, and the \citet{martis2019} measurements from \texttt{MAGPHYS} for galaxies in UltraVISTA at $z = 3-4$.   \citet{Leja2019}  use Bayesian modeling methods with flexible star-formation histories (SFHs) that allow for more accurate estimates of galaxy SFRs (derived from panchromatic data, i.e. UV $-$ mid-IR).  These estimates are appropriate for galaxies out to $z\lesssim 3$ as they are based on galaxies selected in the $H$-band (and at higher redshifts, the Balmer/4000~\AA\ break shifts beyond the \hst/WFC3 $H$-band coverage).   The catalogs of \citet{martis2019} are based on UV $-$ far-IR SED modeling using a rising SFH that accounts for stochasticity. For massive galaxies ($\log M_*/M_\odot > 10.5$) these provide useful constraints on distinguishing star-forming and quiescent galaxies at $z > 3$.  \editone{Both catalogs include the effects of emission lines on the rest-frame colors, which allows us to also test contamination from sources with strong emission lines.} We described these catalogs in detail in Section \ref{section:data}. 

We then define regions in the specific SFR (sSFR)--stellar mass plane that separates star-forming galaxies from those that are quiescent.  We identify a ``ridgeline'' i.e. the density peak in the SFR-mass plane, which corresponds to star-forming galaxies (\citealt{renzini&peng2015}).  At lower sSFRs lies the ``green valley'', below which lie the quiescent galaxies (\citealt{schawinski2014}, \citealt{sherman2021}). The advantage of this approach is that it does not depend on any pre-separation of quiescent and star-forming galaxies, because the star--forming sequence is directly measured from the data. 

We then divide the data into five redshift bins from $0.5 < z < 4$ and determine its 3D distribution by dividing the data into $0.3$ dex square bins. In each redshift bin, we define the green valley by determining where the data falls below 10\% of the star-forming density peak. In Figure \ref{fig:sfms}, we show the two- and three-dimensional SFR-M* plane as well as the green valley, denoted by the white line. The mass range covered by the data reflects the $90\%$ completeness limit at each redshift (taken from \citealt{skelton2014} for \threedhst\ and \citealt{martis2019} and \citealt{marsan2022} for UltraVISTA). \uvj-quiescent galaxies are shown in red. We define ``true'' quiescent galaxies as those with a sSFR below the estimated location of the ``green valley'' using the definition above. 

\subsubsection{Results at z $\leq$ 4}\label{subsection:z4results} 
\begin{figure*}
\begin{center}
\hspace*{-0.2cm} 
\includegraphics[trim={5cm 2cm 5cm 2cm},clip, scale=0.78, angle=-90]{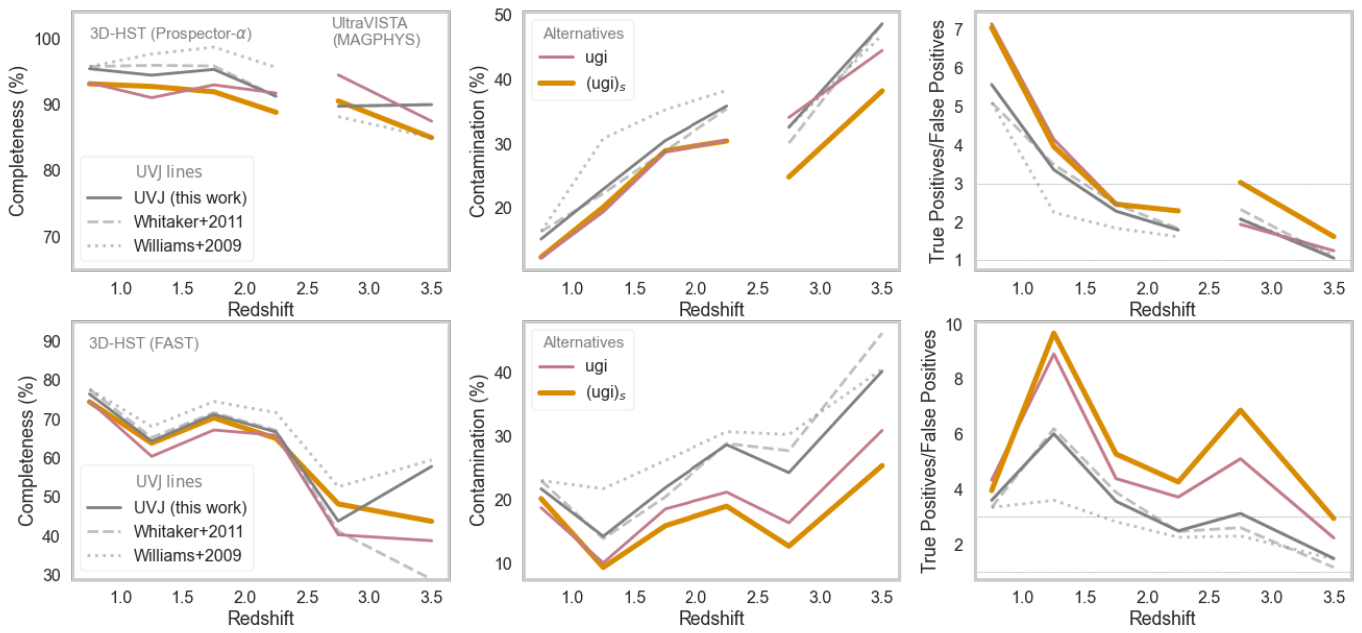}
\caption{Purity tests of color-color selected quiescent galaxies.  The \emph{\bf Top} row shows results using \pa\ fits to the 3D-HST catalogs (\citealt{Leja2019}) for galaxies at $z < 2.5$ and the UltraVISTA/\texttt{MAGPHYS} catalog (\citealt{martis2019}) for galaxies at $z > 2.5$.   The {\emph{\bf Bottom }} row shows results using SFRs and stellar masses from FAST, which is the traditional choice in the literature.  In each row, the columns show the completeness (\emph{\bf Left}), contamination (\emph{\bf Middle}), and true-to-false positive ratio (TP/FP) for quiescent galaxies selected by three color-color methods ($UVJ$, $ugi$, and \sugi, as labeled). In all cases we define true quiescent galaxies based on specific SFRs (i.e. by defining the star-forming main sequence in each redshift bin; Section \ref{section:purity}). In all cases we find that the different methods produce similar completeness.  However, the contamination of \sugi\ is up to $\sim 15\%$ less than that of the other color-color selections.  For this reason the TP/FP ratio is highest for \sugi.  For example, the TP/FP rate for $UVJ$ selection dwindles to $\sim 1$ by $z = 3.5$ (where there is an equal number of false positives for each true positive detected).  In contrast, \sugi-selected samples have a TP/FP ratio that is nearly a factor of 2 higher.}
\label{fig:purity}
\end{center}
\end{figure*}

\editone{In Figures \ref{fig:demographics} and \ref{fig:purity}, we contrast the performance of the \sugi\ diagram with those of \ugi\ and \uvj. Figure \ref{fig:demographics} shows that our definition of quiescence correctly selects galaxies with low sSFRs and low dust extinction (as a function of redshift), and that majority of the contaminants are dusty galaxies with low sSFRs. In Figure \ref{fig:purity}, we quantify this performance, also showing results using \uvj-selection criteria from \citet{williams2009} and \citet{whitaker2011}.} Completeness here is defined as the number of true positives (TP; galaxies with specific SFRs below the threshold of the green valley in Figure \ref{fig:sfms}) in the quiescent region divided  by the total number of true positives in the sample. \editone{By this definition, the false negative fraction (i.e. the number of true quiescent galaxies missed) is 1 $-$ Completeness}. The contamination rate is the number of selected false positives (FP; galaxies that are selected to be quiescent, but have specific SFRs above the green-valley threshold in Figure~\ref{fig:sfms}) divided by the total number of selected galaxies. \editone{Thus, the sample purity is $1-$Contamination.} We also evaluate the ratio of true-to-false positives (TP/FP) as an additional criterion of efficiency.  Note that our definitions of completeness and contamination have different denominators. This is why the contamination rate in Figure \ref{fig:purity} is not $50\%$ at $z = 3$ even though the \uvj\ TP/FP ratio is 2. 

We see that the \sugi\ diagram, although comparable in completeness to \uvj, has significantly less contamination, hence a much higher TP/FP ratio as a function of redshift. A slightly smaller wavelength baseline in $(u-g)$ and $(g-i)$ means less quiescent galaxies overall are selected in \sugi\ and \ugi\ (hence \sugi\ is $\sim 5\%$ less complete) at lower redshifts. At higher redshifts ($z > 2.5$), we start to see the effects of extrapolation and increased contamination from strong \oiii\ emitters on \uvj\ and \ugi, problems which the \sugi\ filters are designed to fix. By $z = 3.5$, the \uvj\ selection has a TP/FP ratio of $\sim 1$, which means that \emph{the contaminants in the \uvj-selected quiescent sample are just as numerous as the true quiescent galaxies}.

SFRs and stellar masses from SED-fitting codes have a  persistent 0.5 and 0.2 dex scatter (see Section \ref{section:3dhst}), respectively. While the flexible SFHs and and self-consistent SED modeling employed in \pa\ and \texttt{MAGPHYS} reduce systematics and hence provide an improved ability to mitigate this effect, it is still worth noting. In order to investigate the effect of this offset on our derived completeness and contamination fractions, we repeat our analysis above using SFRs and stellar masses in the 3D-HST catalogs derived using \texttt{FAST}, which uses more a simplistic parameterization of the SFH (usually exponential or "delayed exponential" models) and tends to be the traditional choice for estimating stellar population parameters in the literature. Figure~\ref{fig:purity} (bottom panel) shows the results from these tests. All three methods (\uvj, \ugi, \sugi) using \texttt{FAST} have lower completeness rates. We attribute this to the adopted star formation history, which has been shown to produce highly biased stellar population parameters (e.g. \citealt{carnall2019}), favoring young stellar populations, hence higher sSFRs. This particularly impacts red galaxies (which could be either dusty star-forming galaxies or quiescent galaxies) at high-redshift, for which this SFH has been shown to be inadequate (see Section \ref{section:uvista}). Nevertheless, the effect appears to be systematic, affecting all color-color methods similarly. 
\begin{figure*}
\begin{center}
\includegraphics[trim={2cm 4cm 0cm 3cm},clip,scale=0.45]{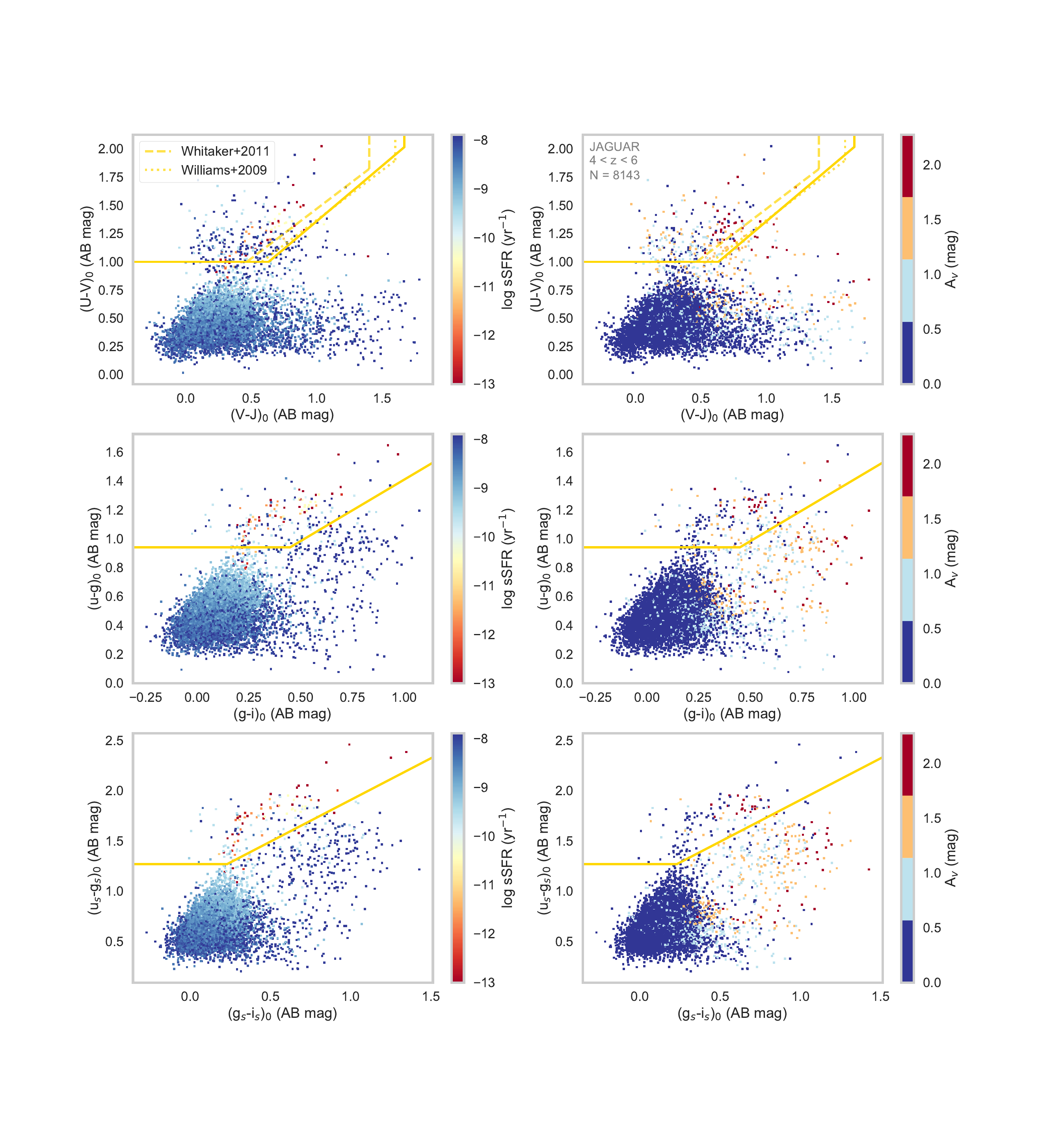}
\caption{Color-color diagrams for simulated data with added noise at $4 < z < 6$. Each row shows color-color diagrams for galaxies with log(M$_*$/M$_\odot$) $\geq$ 8.9 in the JAGUAR catalog, color-coded by specific star formation rate (left panel in each row) and dust extinction (right panel in each row).  The \emph{\bf Top} row shows the \uvj\ colors.  The \emph{\bf Middle} row shows the \ugi\ colors.  The \emph{\bf Bottom} row shows the \sugi\ colors.  In each case, the  rest-frame colors were derived using $HST$, \jwst/NIRCam and {\it Spitzer}/IRAC photometry from 0.4-5$\mu m$. The ($\rm U-V$)$_0$, ($u-g$)$_0$, and ($u_s-g_s)_0$ horizontal lines were each lowered by 0.23 mag to accommodate the increasingly younger population of quiescent galaxies at these redshifts, although we maintained the slopes and zeropoints from Figure \ref{fig:calibrations}.}
\label{fig:jaguar_cc}
\end{center}
\end{figure*}
Our results with \texttt{FAST} mirror those derived from \pa\ and \texttt{MAGPHYS} above: the completeness of quiescent galaxies is similar for all three color-color methods (left column of Figure~\ref{fig:purity}). However, the contamination in the quiescent selection is consistently lower using the  \sugi\ colors at $z > 3$.  Finally, the \sugi\ color-selection of quiescent galaxies has the highest TP/FP ratios (right column of Fig~\ref{fig:purity}).  This is primarily a result of the lower contamination from the \sugi\ selection, and for this reason we see the largest improvement in TP/FP at $z > 3$.   This fact will be crucial at these redshifts as contaminants (false positives) can otherwise exceed true positives, as in \uvj. 

Our results for the completeness and contamination of $UVJ$-selected quiescent galaxies at $z \leq 4$ are broadly consistent with similar results in the literature using photometric data from large area surveys, spectroscopic campaigns, and simulations. Spectroscopic studies at $z <$ 1 have shown that \uvj-selected quiescent galaxies at various mass cutoffs show $\sim 10-65\%$ contamination from galaxies with signs of ongoing star formation (e.g. \citealt{tacchella2022}). Using a sample of log (M$_*$/M$_\odot$) $>$ 10 galaxies at $1.5 <z < 2.5$ in 3D-HST, \citet{leja&tacchella&conroy2019} find that galaxies with the lowest specific-SFRs (log(sSFR) $<$ -10.5) show up to $\sim 30\%$ contamination. Using CANDELS, \citet{shahidi2020} find 80\% completeness for log(M$_*$/M$_\odot) > 10$ galaxies at $2.8 < z < 4$ and 50\% completeness at $z  = 4-5.4$. Using SHELA (\citealt{papovich2016}, \citealt{wold2019}), \citet{sherman2021} find up to 38\% contamination from dusty galaxies for log(M$_*$/M$_\odot$) $>11$ quiescent galaxies at $1.5 < z < 3$. Finally, \citet{diaz-garcia2019} find a 20\% contamination rate at $z \sim 1$ from dusty galaxies using data from the ALHAMBRA survey (\citealt{moles2008}). Similarly, for the simulations, \citet{lustig2022} find 67-75\% completeness and 50-60\% contamination for log(M$_*$/M$_\odot$) $> 11$ quiescent galaxies at $z = 2.7$ in Magneticum\footnote{\url{www.magneticum.org}} (\citealt{steinborn2016}) and Illustris TNG\footnote{\url{www.tng-project.org}} (e.g. \citealt{springel2018}). 

Lastly, we would like to caution the reader on the importance of calibrating color selection methods on the dataset which they choose to use for their study. A number of studies, including \citet{kawinwanichakij2016} have pointed out that there are systematic offsets in the rest-frame colors of galaxies at fixed mass and redshift in different surveys. This means that quiescent selection lines defined on different datasets will yield different numbers of quiescent galaxies, resulting in different completeness and contamination rates. We demonstrate this by applying the \citet{whitaker2011} and \citet{williams2009} \uvj\ lines to our dataset and reporting their completeness and contamination rates in Figure \ref{fig:purity}. Although the \citet{whitaker2011} line appears to be quite similar to ours, the results for the \citet{williams2009} lines are very different, yielding up to 10\% more quiescent galaxies and contaminants. We also determined during some earlier tests that the \citet{muzzin2013} lines (not shown in Figure \ref{fig:purity}) yield up to 20\% less completeness and 5\% less contamination than the other \uvj\ lines applied to our dataset. For these reasons, although we provide \sugi\ lines, we urge the reader to treat them as initial selections and calibrate the color selection lines on their own dataset in order for the method to be most accurate. 

\subsubsection{Results at z $\geq$ 4: What to expect with JWST}\label{subsection:jaguar}
Using the same methodology and definitions detailed above, we determine the completeness and contamination of the quiescent selections of \uvj, \ugi, and \sugi\ in simulated JWST data using the JAGUAR catalog (\citealt{williams2018}). We do this using a sample of 10,200 galaxies split into two redshift bins: $3 < z < 4$ and $4 < z < 6$. The latter bin is large because there are only a handful of quiescent galaxies in the catalog at these redshifts (TP = 139; Figure \ref{fig:jaguar_cc}). For the $z>4$ bin, we lower the ($U-V$)$_0$ line by 0.23 mag to include more post-starburst galaxies in the quiescent region, because there is evidence from both photometric and spectroscopic samples at high-z that these galaxies tend to have much lower $U-V$ colors than the quenched population (\citealt{marsan2015}, \citealt{carnall2020}, \citealt{marsan2022},\citealt{forrest2020}, \citealt{schreiber2018}, discussed further in Section \ref{section:limitations}). We lowered the horizontal cut by 0.23 mag to match the $U-V$ colors of the young quiescent galaxies in these studies (i.e. ($U-V$)$_0$ $\sim$ 1). In the same vein, we lower the horizontal lines of the other color selection methods (\ugi, \sugi) by the same amount. 

\begin{figure*}
\begin{center}
\includegraphics[trim={0cm 16cm 0cm 0cm},clip,scale=0.27]{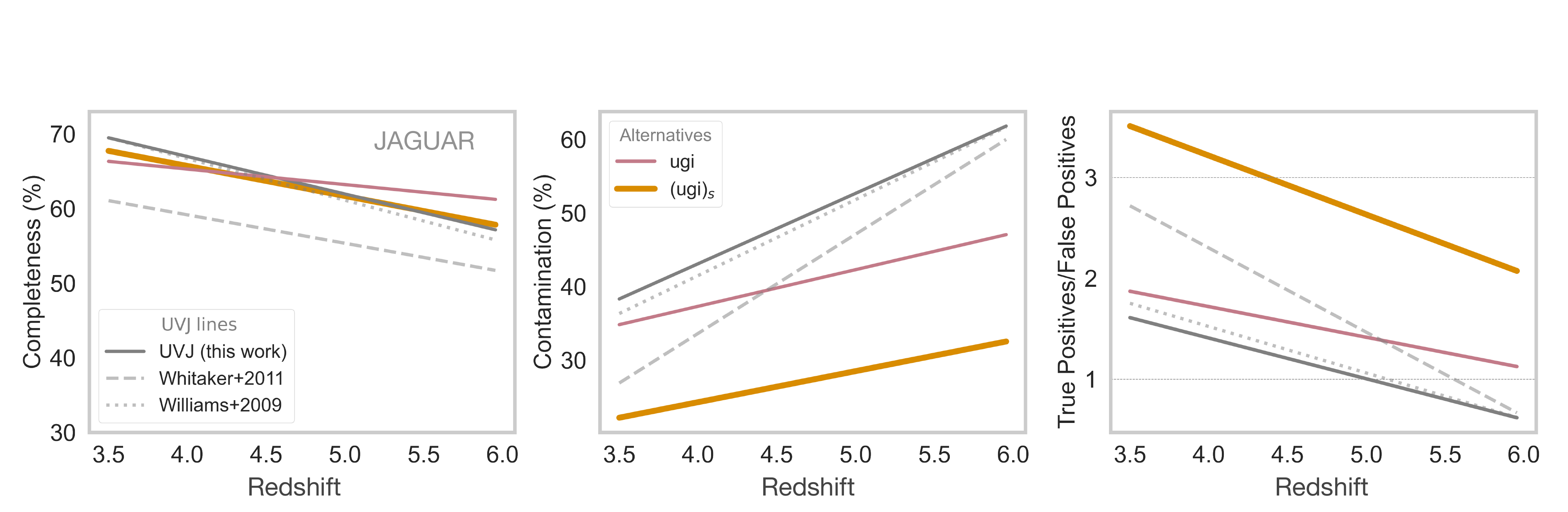}
\caption{Sample selection efficiency of \uvj, \ugi, and \sugi\ at $3 < z < 6$. We show the completeness and contamination of the quiescent selections of \uvj, \ugi, and \sugi\ using specific star formation rates from the JAGUAR catalog, which were generated using \texttt{BEAGLE}. As in Figure \ref{fig:purity}, although the color selection methods are comparable in completeness at these redshifts, \sugi\ has a much lower contamination rate (up to $\sim 30\%$  less) than \uvj\ and \ugi, resulting in a higher true-to-false positive rate at $z = 6$ ($\geq 2.3$). Conversely \ugi\ and \uvj have true-to-false positive rates of $\lesssim 1$ by $z=6$, indicating that the true quiescent galaxies are outnumbered by contaminants, making these methods unfeasible for selecting quiescent galaxies at these redshifts. }
\label{fig:jades_purity}
\end{center}
\end{figure*}

The completeness, contamination, and true-to-false positive rates of of the selection methods, \uvj\, \ugi, \sugi\ are shown as a function of redshift from $z = 3.5$ and $z = 6$ in Figure \ref{fig:jades_purity}.  As in Figure \ref{fig:purity}, although the color selection methods are comparable in completeness at these redshifts, \sugi\ has a much lower contamination rate (up to $\sim 30\%$ less) than \uvj\ and \ugi, resulting in a higher true-to-false positive rate ($\gtrsim 2.5$). Conversely, \ugi\ and \uvj\ have true-to-false positive rates of $\lesssim 1$ by $z = 6$, indicating that the true quiescent galaxies are outnumbered by contaminants, making these methods unfeasible for selecting quiescent galaxies at these redshifts. For this reason, the \sugi-selection method provides improved fidelity in color-selected samples of quiescent galaxies at these high redshifts. It is worth noting that these numbers represent the best case scenario, as the number of contaminant dusty star forming galaxies may be higher in reality, since they tend to be underrepresented in UV-selected samples like JAGUAR (Appendix \ref{section:appendix}).   

\subsection{Contamination from Extreme Emission-Line Galaxies in Quiescent Galaxy Selection}\label{subsection:eelgs}

Here, we test the ability of each color selection method to minimize the number of galaxies with strong emission lines in the quiescent region. Extreme emission line galaxies (EELGs) tend to have EW$_0$ $\sim$ 100 Å $-$ 1000 $\AA$ in \oii\ $\lambda \lambda$ 3726, 3729, \hb +\oiii\, and H$\alpha$+\nii\, although a handful of massive galaxies have been discovered at $z >$ 7 with EW(\hb+\oiii) $\approx$ 1000$-$2000 $\AA$ (\citealt{smit2014, smit2015}, \citealt{roberts-borsani2016}, \citealt{castellano2017}). At $z\sim 0.1$, EELGs correspond to rare, starbursting, low-mass galaxies (log(M$_*$/M$_\odot$) $\sim8 -9$) where the mass doubling time can be 100 Myr (e.g., \citealt{izotov2016}). These galaxies are 10$-100\times$ more common at $z \gtrsim1$ (\citealt{vanderWel2011}, \citealt{maseda2018}) than at $z \leq 0.5$ and may be the dominant star-formation mode at $z  = 3-4$ (\citealt{kurczynski2016}, \citealt{cohn2018}, \citealt{tran2020}). 

Galaxies with strong emission lines can mimic the rest-frame \uvj\ colors of galaxies with old stellar populations by boosting their photometry and making them appear redder in ($U-V$)$_0$ than they actually are (e.g. \citealt{labbe2005}). Strong emission lines can be a result of either ongoing star formation or AGN activity, both of which are observed to increase at higher redshifts (e.g. \citealt{sobral2014}, \citealt{d'eugenio2020}, \citealt{marsan2022}). \hb+\oiii\ emission with EW $\leq$ 200 $\AA$ has been shown to boost fluxes in the $K$-band\footnote{Originally shown by \citealt{shapley2005}.} by 5-30\% (\citealt{schreiber2018}, \citealt{forrest2020_survey}) and in the IRAC bands by up to 65\% (\citealt{labbe2013}). Based on their central wavelengths and widths, the synthetic \su\ and \si\ filters require extremely strong lines (\editone{rest-frame} EW $>$ 800 $\AA$ in \oii\ and H$\alpha$+\nii\, respectively) to be contaminated. The synthetic \sg\ filter, however, has only a 163 $\AA$ buffer from the central wavelength of \hb$\lambda 4861$. This implies that it can be contaminated by galaxies with \hb+\oiii\ $> $500 $\AA$. While the fraction of star-forming galaxies with \hb+\oiii\ EW $>$ 500 $\AA$ is low at $z =1.7- 2.3$ ($5.4-8.2$\%, \citealt{boyett2022}), it increases to $\sim 40\%$ at $z = 5$ (\citealt{rasappu2016}\footnote{converted from EW(H$\alpha$) by \citealt{boyett2022}}) and could reach $\sim 50\%$ at $z = 6$, assuming that the rest-frame optical emission evolves as a power law, (1+z)$^P$, and given that fractions at $z \sim7$ are measured to be $\sim 65\%$ (\citealt{endsley2021}). Consequently, we can expect an increasingly larger number of EELGs to contaminate the quiescent region at higher redshifts by boosting the fluxes in the reddest observed photometric bands ($K$-band $-$ IRAC channel 4).

\begin{figure*}
\begin{center}
\includegraphics[trim={7cm 1cm 2cm 2cm},clip,scale=0.27]{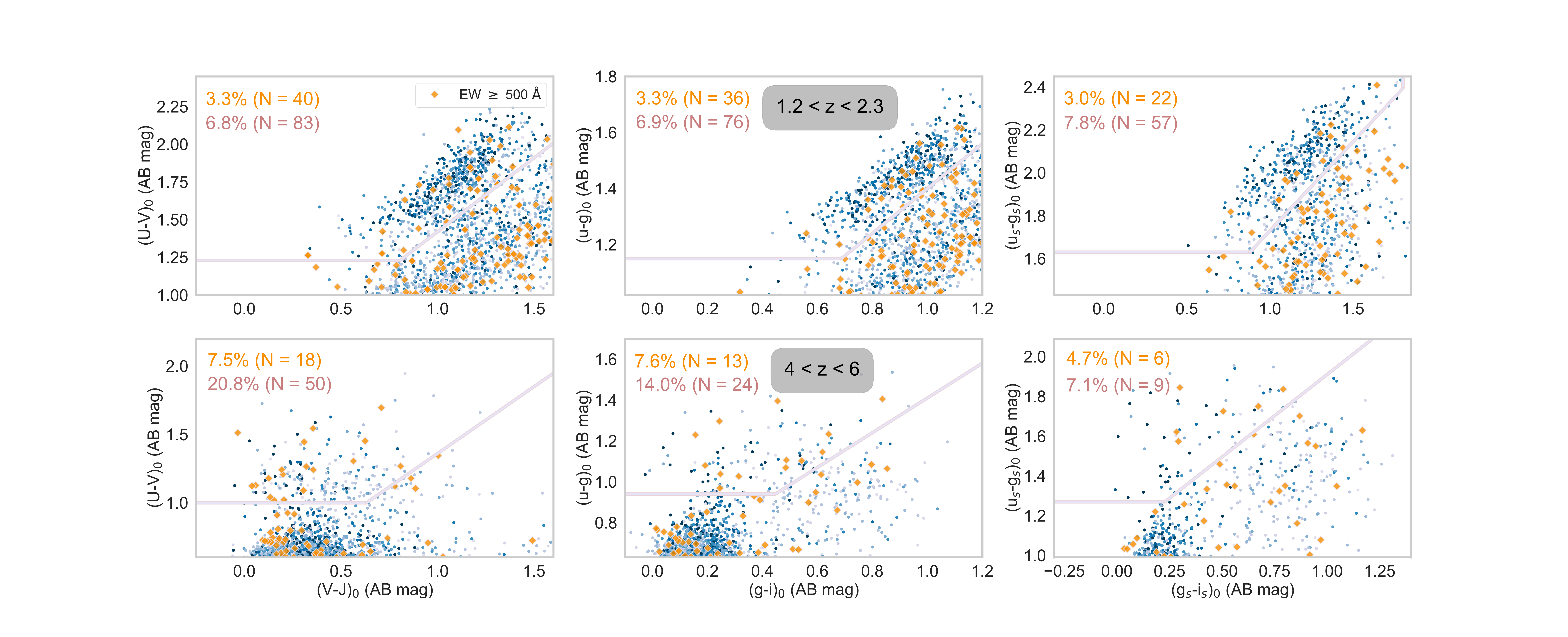}
\caption{Fraction of extreme emission line galaxies in \uvj, \ugi, and \sugi\ using mass-complete samples from \threedhst\ and JAGUAR at $1.2 < z < 2.3$ (\emph{\bf Top}) and $4 < z < 6$ (\emph {\bf Bottom}), respectively. For each redshift bin, each color diagram shares the same line parameters as Figure \ref{fig:jaguar_cc}. As in Figure \ref{fig:jaguar_cc}, the ordinate limits and horizontal color lines shown are different for each diagram in order to capture the evolution of galaxies in color space as a function of redshift. The fractions and numbers of \editone{rest-frame} EW $>$ 500 $\AA$ galaxies selected as quiescent in each color diagram are shown in orange. For the EW $>$ 500 $\AA$ population in both redshift bins,  although the methods have similar fractions of EELGs in their quiescent regions, the \sugi\ diagram has a smaller number of them (a factor of 1.6 lower). \editone{We also show the numbers and fractions of EW $>$ 300 $\AA$ in brown (not plotted) for comparison.} For the EW$ >300$ $\AA$ EELG population, the contamination in \sugi\ is lower by a factor of 3. }
\label{fig:high_ew}
\end{center}
\end{figure*}

We quantify this contamination for \uvj, \ugi\, and \sugi\ using mass-complete samples from \threedhst\ and JAGUAR at $1.2 < z < 2.3$ and $4 < z < 6$, respectively. The \threedhst\ line measurements are from \hst/WFC3 G141 grism observations (\citealt{momcheva2016}) and those from the simulated JAGUAR catalog are determined using the \texttt{BEAGLE} SED fitting code (\citealt{williams2018}). In Figure \ref{fig:high_ew}, we show the fraction of galaxies in the quiescent region with \hb+\oiii\ equivalent widths $>$500 $\AA$ for all three color selection methods. For the EW $>$ 500 $\AA$ population at both redshift ranges,  although the methods have similar fractions of EELGs in their quiescent regions, the \sugi\ diagram has a smaller number of them (a factor of $1.6$ lower). However, for the EW$>300$ $\AA$ EELG population, the contamination in \sugi\ is lower by a factor of 3. For comparison, EW $>$ 200 $\AA$ contamination fractions at $z >$ 4 are: 0.38 (\uvj, N=92), 0.22 (\ugi, N = 37), 0.11 (\sugi, N = 14). \editone{Photometric redshift uncertainties could also be a factor (as an erroneous redshift could shift an emission line into the rest-frame band, though this effect will be lessened for galaxies with stronger emission lines as these tend to have lower redshift uncertainties, see \citealt{momcheva2016}).  Nevertheless, in  Appendix \ref{section:appendix_b} we show that photometric redshift errors of $\sigma_z/(1+z) < 9.1$\% correspond to fewer than 16\% of sources having their rest-frame colors impacted by emission lines (and most modern surveys have better photometric redshift accuracy than this). Therefore we do not expect this to significantly impact the rest-frame colors.  } Our results are consistent with spectroscopic surveys, which report a contamination rate of 21-30\% from EW(\hb+\oiii) $>$ 100 $\AA$ galaxies in \uvj-selected quiescent samples at $z = 3-4$, with at least a third of those having f$_{\hbox{[\ion{O}{3}]}}$/f$_{\rm H\beta} >$ 6 (\citealt{schreiber2018}, \citealt{forrest2020_survey}),  indicative of AGN activity, particularly at high mass (e.g. \citealt{strom2017}, \citealt{belli2017}, \citealt{reddy2018}). \edittwo{In Appendix \ref{section:appendix_c}, we show that \sugi\ screens all of the star-forming contaminants in \uvj\ and \ugi\ at $3 < z < 4$.}  

\subsection{Implications for Quiescent Galaxy Selection at $z > 4$:  Current and Future}
\subsubsection{The Beginning of the End}\label{subsection:beginning_end}
We are approaching the end of the era of discovery of quiescent galaxies. We have spent the past two decades attempting to obtain a complete census of different populations of galaxies as a function of redshift, and color selection methods have played a pivotal role in this (e.g. \citealt{daddi2004}, \citealt{kriek2009}, \citealt{quadri2007}, \citealt{guo2013}, \citealt{nayyeri2014}). Color selection methods are useful because they can quickly isolate different galaxy populations in large datasets over a range of redshifts using only 3 bands.  Although star-forming galaxies may well exist out to $z \sim 13$ (\citealt{harikane2022}), we may be reaching the era where there are no more quiescent galaxies to be selected ($z\sim6$). We have spectroscopically confirmed quiescent galaxies out to $z \sim 4$ (\citealt{mccarthy2004}, \citealt{cimatti2004}, \citealt{marsan2015}, \citealt{valentino2020}, \citealt{forrest2020}, \citealt{saracco2020}, \citealt{tanaka2019}, \citealt{glazebrook2017}) and we have a handful of photometrically-selected candidates at $z = 5-6$ (e.g. \citealt{marsan2021}, \citealt{merlin2019}, \citealt{mawatari2016}) with searches underway using ground-based surveys that leverage the detection capabilities of medium-band filters  (\citealt{esdaile2021}). However, given that passive fractions are $<10\%$ at $z>4$ (e.g. \citealt{santini2021}), by $z\sim6$ we will be witnessing the emergence of the first quiescent galaxies. These quiescent galaxies will be some of the first to have started forming stars and quench and will be important for answering questions about structure formation in the early universe and settling debates about quenching mechanisms (see \citealt{nanayakkara2022} for a review). With this in mind, it is ever so important that we select pure and complete samples of quiescent galaxies for spectroscopic follow-up so that we do not miss any of this interesting population. 

\begin{figure}
\includegraphics[trim={0cm 0cm 0cm 0cm},clip,scale=0.48]{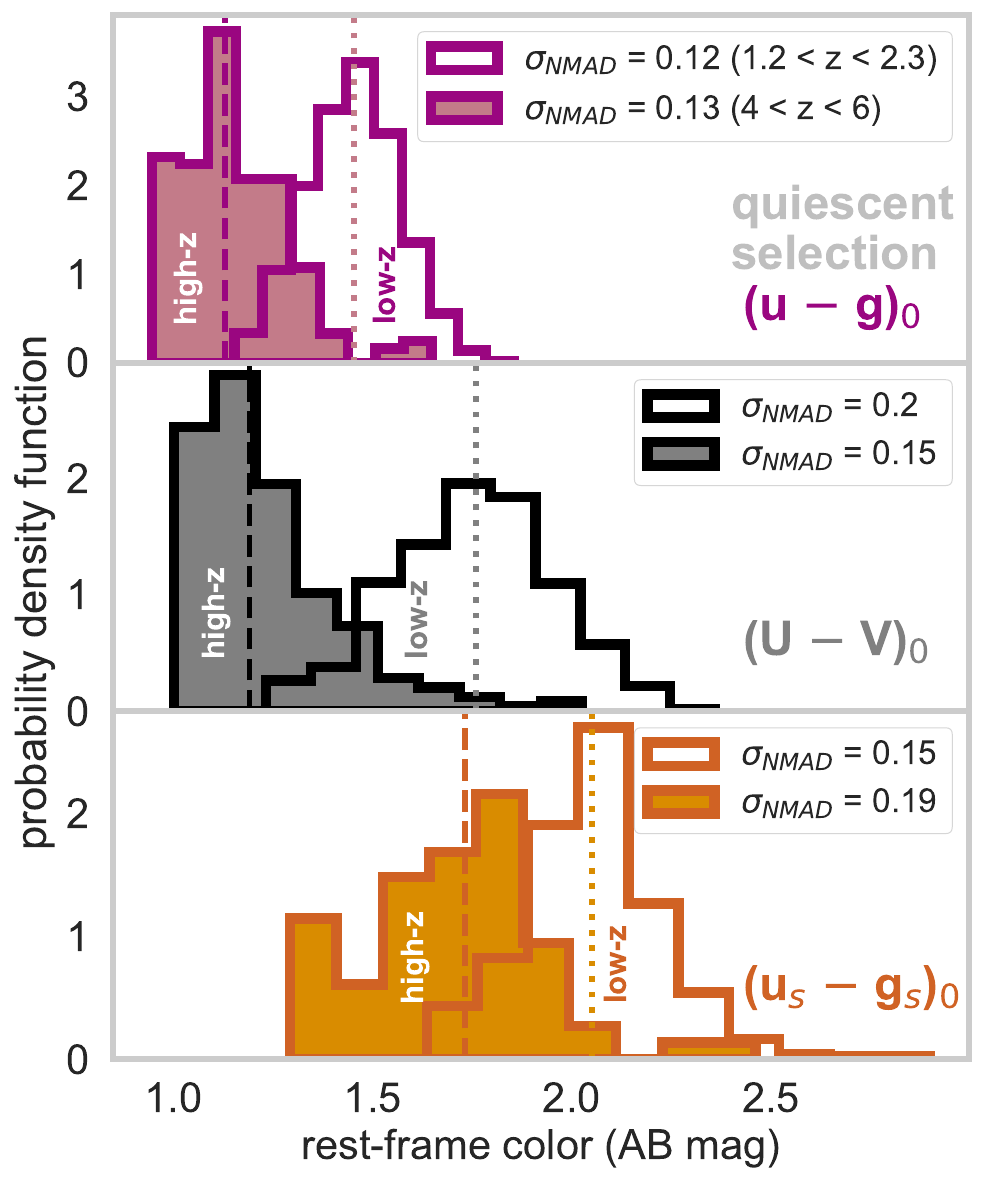}
\caption{Vertical color of the red sequence in each color selection method at low and high redshift (1.2 $< z <$ 2.3 and 4 $< z <$ 6) for \ugi\ ({\bf Top}), \uvj\ ({\bf Middle}), and \sugi\ ({\bf Bottom}).  The associated median and normalized median absolute deviation for each distribution are also shown. Galaxies at $z<$ 2.3 (low-z) tend to exhibit redder colors than those at $z>4$ (high-z) because the strength of the Balmer/4000 $\AA$ break decreases as a function of redshift. The (\su$-$\sg)$_0$ color, however, is reddest in both bins, with a median of 1.73 mag (2.05) at $z >4$ ($z<2.3$), compared with 1.19 mag (1.76) in ($U-V$)$_0$ and 1.13 (1.45) mag in ($u-g$)$_0$. \sugi\ also has a larger spread in colors. Both of these indicate that \sugi\ is better at isolating quiescent galaxies and capturing the diversity of quiescent SEDs in the high-z Universe.}
\label{fig:color_dist}
\end{figure}

\begin{figure}
\includegraphics[trim={1.5cm 2cm 1.5cm 2cm},clip,scale=0.38]{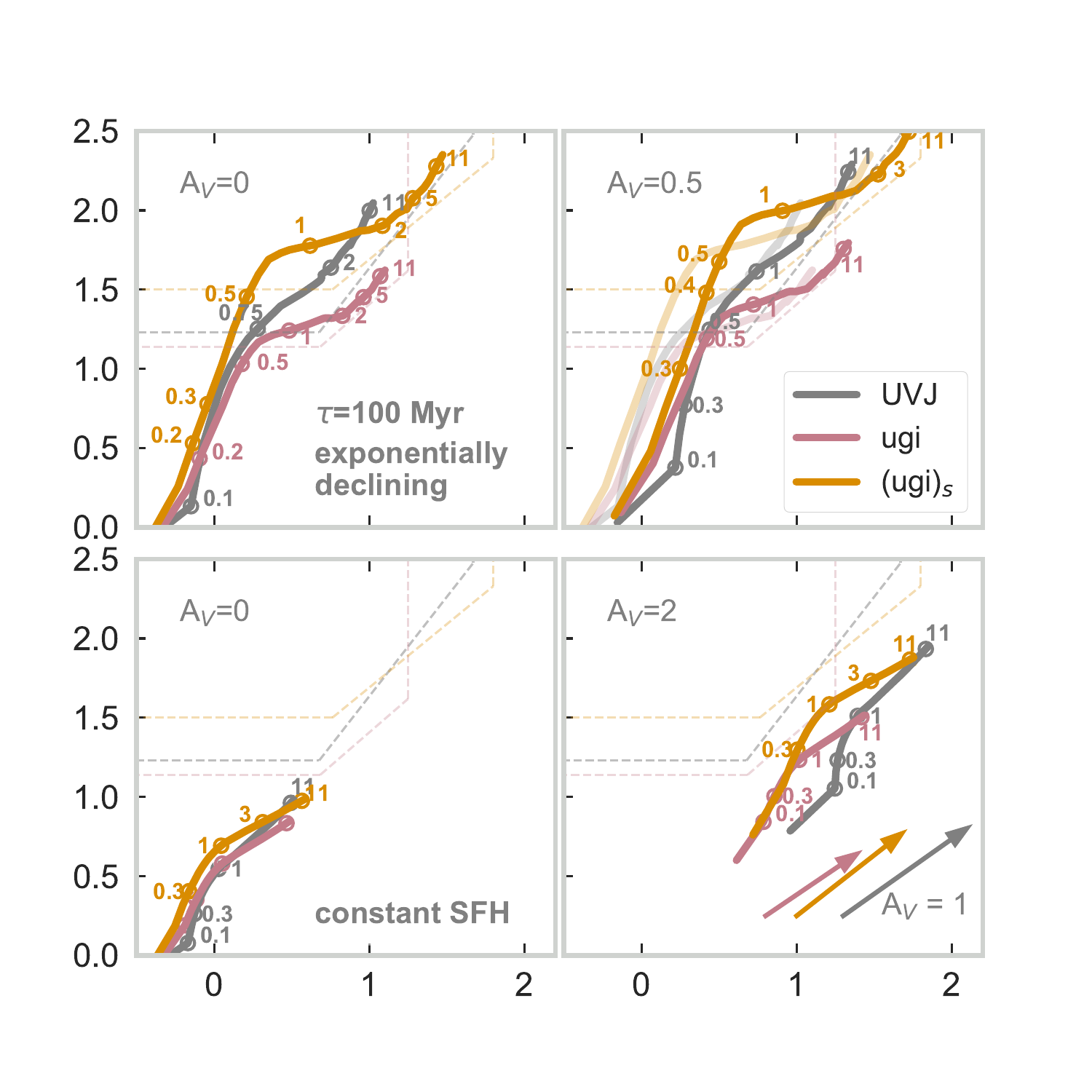}
\caption{\edittwo{Superimposed color evolution tracks from Figure \ref{fig:sps_tracks} for \uvj, \ugi, and \sugi. {\bf Top}: tracks for an exponentially-declining SFH with e-folding time $\tau = 100$ Myr with no dust (\emph{Left}) and A$_{\rm V}$ = $0.5$ (\emph{Right}). We also show the dust-free tracks in this panel for reference. {\bf Bottom}: Constant SFH with no dust (\emph{Left}) and A$_{\rm V}$ $= 2$ (\emph{Right}). All tracks assume a \cite{calzetti2000} dust extinction curve. {\bf Top Left}: The $\tau = 100$ Myr stellar population track evolves into the quiescent region at younger ages in \sugi\ than in \uvj\ and \ugi\ ($\sim 250$ and 150 Myr younger, respectively). {\bf Top Right}: Moderate dust extinction causes the color tracks to enter the quiescent region at younger ages in all three methods. While this would cause star-forming galaxies to enter the quiescent sample, \sugi\ is better at reducing contamination from galaxies with strong emission lines. This is likely why it selects fewer moderately-dusty galaxies (see Figure \ref{fig:demographics})}.}
\label{fig:sps_tracks_comb}
\end{figure}

In this \emph{Paper}, we have presented three synthetic filters: \su\, \sg, and \si\, which accomplish this goal most efficiently, using data from our largest and deepest ground-based photometric surveys and simulated {\emph {JWST}} data. We consider one of our intended goals in designing our filter set: achieving increased sensitivity to the Balmer/4000 $\AA$ break. In Figure \ref{fig:color_dist}, we show the distribution of the vertical colors of galaxies in the quiescent regions of \uvj, \ugi\, and \sugi\ at $1.2 < z < 2.3$ and $4 < z < 6$. The strength of the Balmer break is predicted to be lower  at fixed sSFR for galaxies at $z = 2-10$ than for their lower redshift counterparts (\citealt{shen2020}). If so, we expect the vertical colors in the high-z bin to be bluer than those in the low-z bin and that the most efficient color selection method would have redder vertical colors overall, signaling that the Balmer/4000 $\AA$ break produces a stronger color signature in that color space. This is precisely what we see in Figure \ref{fig:color_dist}, with median values of (\su$-$\sg)$_0$ = 1.73, ($U-V$)$_0$ = 1.19, and ($u-g$)$_0$ = 1.13 at $z >$ 4. A wider range in (\su$-$\sg)$_0$ ($\sigma_{NMAD}$ = 0.19) over ($U-V$)$_0$ (0.15) and ($u-g$)$_0$ (0.13) at those redshifts may also indicate that \sugi\ is better at capturing the diversity of quiescent SEDs at higher redshift than \uvj\ and \ugi. \edittwo{Indeed, galaxies enter the quiescent region at younger ages in \sugi\ than in \uvj\ and \ugi\ ($\sim$ 250 and 150 Myr younger, respectively in Figure \ref{fig:sps_tracks_comb}), which allows \sugi\ to select young post-starbursts typically missed in \uvj\ selection (Figures \ref{fig:missed_objects_sugi_uvj} and \ref{fig:uvj_missed_seds}).} This property will be extremely useful for studying the age gradients of the first quiescent galaxies (e.g. \citealt{whitaker2013}). 

This is particularly important because so far, most of the quiescent galaxies that have been confirmed spectroscopically at $z >$ 3 are young, recently-quenched galaxies, with a few exceptions (e.g. Galaxy D in \citealt{kalita2021}). Older quiescent galaxies are predicted to be $1-2$ magnitudes fainter in the $K$-band than their post-starburst counterparts at the same redshift, with stronger Balmer/4000 $\AA$ breaks and weaker absorption lines due to their evolved stellar populations (\citealt{forrest2020}). This suggests that their absence from current surveys is simply a selection effect, therefore this population should be revealed with deep ongoing NIR surveys e.g. FENIKS (\citealt{esdaile2021}) and the upcoming {\emph{JWST}} ERS, GTO, and GO programs, which will enable us to detect some of the faintest galaxies in the distant universe. 

\subsubsection{Possible Limitations of the \sugi\ Diagram}\label{section:limitations}
Next, we consider an apparent limitation of \sugi, and color-color-selection methods in general at high-z. If indeed \sugi\ is more efficient at selecting high-z quiescent galaxies, why then is it $\lesssim$ 70\% complete at $z>4$ (Figure \ref{fig:jades_purity})? First, the simplest explanation is that they were scattered out of the quiescent region due to the addition of noise, since that would preferentially affect faint and low mass galaxies. In fact, the completeness of quiescent galaxies is reduced by $\sim 20\%$ for all three methods after we add noise (Appendix \ref{section:appendix}). 

A second possibility is that we are missing some quiescent galaxies because our adopted slope for the \sugi\ diagonal line excludes them. For the sake of consistency, we assumed that the slope of the diagonal line remains constant as a function of redshift, similar to what others have done (\citealt{williams2009}, \citealt{muzzin2013}, \citealt{whitaker2011}). Like those studies, we lowered the horizontal cut to account for the evolution in color due to quiescent galaxies having increasingly younger ages at higher redshifts (see discussion in paragraph below). However, it is not entirely clear if the slope needs to evolve as well, particularly at $z>$3. This can be resolved with larger unbiased samples of quiescent galaxies with high signal-to-noise photometry. 

A third possibility is that lowering the horizontal lines of each color diagram by 0.23 mag is not enough to capture all the post-starbursts in the sample. This would be the case for dust-free star-forming galaxies that quenched abruptly ($< 175$ Myr) prior to observation, as they can have ($U-V$)$_0$ colors as low as 0.5 mag (\citealt{merlin2018_1}), which suggests that genuinely quiescent galaxies can be found outside of the fiducial \uvj\ boundaries, and by extension, the \sugi\ boundaries as well. \edittwo{For example, the 320 Myr old post-starburst galaxy in \citealt{forrest2020_survey} is missed in both \uvj\ and \sugi\ (Figures \ref{fig:missed_demographics} and \ref{fig:missed_seds})}. Metal-poor galaxies ($Z = 0.02\; Z_\odot$ ), which take $\sim3$ Gyr to become \uvj-quiescent, also meet this fate, as the time needed to enter the quiescent region exceeds the age of the Universe at that redshift (\citealt{tacchella2018}). Post-starburst galaxies are predicted to comprise an increasingly larger fraction of the quiescent population at $z>$3 because galaxies will have less time to evolve passively after quenching.  One possible solution is to use a slightly redder \su\ filter ($\lambda_c = 3200\; \AA$, as opposed to 2900 $\AA$) at $z>4$. This would produce a stronger (\su$-$\sg)$_0$ color signature for the entire quiescent population, potentially moving more post-starbursts into the quiescent region. This, however, may be at the risk of increasing the contamination from dusty star-forming galaxies, as this modification would reduce the wavelength baseline needed to accurately measure dust reddening in the near-infrared. \editone{Others have resolved this by eliminating the horizontal ($U-V$) cut altogether and instead extending the diagonal line to bluer colors (e.g. \citealt{marsan2022}, \citealt{forrest2020_survey}). While this would increase the number of post-starbursts in the quiescent region, it would also unfortunately, increase the number of non-dusty star-forming contaminants. It is also important to note that the way the slope of the diagonal line is determined becomes even more critical in this case (see our cautionary note in Section \ref{subsection:z4results}.)}

The fourth and final possible reason for $<70\%$ completeness at $z>4$ is that we may be missing old galaxies that are significantly obscured by dust (E(B-V) $\gtrsim$ 0.4, i.e. A$_{\rm V}$ $\gtrsim 1$), causing them to fail the $V-J$ (hence the ($\sg-\si$)$_0$) cut (\citealt{merlin2018_1}). \editone{Although local ($z \sim 0$) massive, quiescent galaxies have little-to-no dust \citep{smith2012}, recent observations have shown that their quiescent counterparts at higher redshifts ($z \gtrsim 1.5$) could be dusty, with up to $12\%$ showing MIPS and \emph{Herschel} detections at $z = 3 - 4$ \citep{martis2019}. }This estimate does not take into consideration gas depletion timescales, which are reported to be quite short (100$-$600 Myr) for quiescent galaxies at $z \sim 1.5-3$ (\citealt{williams2021}, \citealt{caliendo2021}, \citealt{whitaker2021}). It may still be possible to catch a quiescent galaxy in its dusty phase if the quenching timescale is longer than the gas depletion timescale. A few galaxies with some measure of dust obscuration (verified via ALMA detections) have been reported in the literature (\citealt{schreiber2018}, \citealt{whitaker_nature2021}) and found in simulations (\citealt{akins2022}). While it is possible that these galaxies may indeed represent an evolutionary phase for quiescent galaxies, it is equally likely that they are dusty star-forming galaxies masquerading as quiescent galaxies due to their low sSFRs (heavily-dependent on the choice of SFH, e.g. \citealt{marchesini2010}). 

We can resolve this through targeted spectroscopic campaigns searching for the presence (or lack of) absorption features, which would unambiguously confirm their quiescent nature. Additionally, flexible SFHs and self-consistent SED modeling are recommended because they have been shown to better approximate stellar masses as they account for emission from old stars (\citealt{martis2019}, \citealt{Leja2019}), thereby resulting in more accurate sSFRs for dusty star-forming galaxies and lower dust masses for quiescent galaxies (\citealt{whitaker_nature2021}). In summary, if this population of high-z dusty quiescent galaxies is real, removing the vertical (\sg-\si) color cut (e.g. \citealt{martis2019}, \citealt{marsan2022}) would increase completeness, but would also increase contamination from dusty star-forming galaxies.

\edittwo{It is also important to note that completeness rates are subject to how quiescence is defined (i.e. the estimated location of the SFMS and green valley, see Section \ref{subsection:completeness_contamination} for a thorough discussion). We find that \emph{none} of the low sSFR galaxies (sSFR $<$ SFMS $- 0.7$) at $3 < z < 4$ that were missed in \sugi\ are truly quiescent (Figures \ref{fig:missed_demographics} and \ref{fig:missed_seds}). This means that we have underestimated our completeness rates at $3 < z < 4$ and may be $>70\%$ complete at $z > 4$. One possible solution to this is to consider the M$_*$ $-$SFR uncertainties for each galaxy when estimating the star forming density peak (e.g. \cite{leja2022}). This is particularly important for the high-mass end, where 1) a large fraction of galaxies have relatively low SFRs (hence higher SFR uncertainties) and 2) there are few galaxies in total (so the the uncertainties become even more important). Another solution is to consider using a double power-law fit to the SFMS, which allows more flexibility at high masses.} 

To conclude, we consider the utility and accuracy of color selection methods compared to others in the literature. Are there better ones out there? The short answer is yes. The longer answer is no. Other methods such as template fitting (e.g. \citealt{shahidi2020}) and unsupervised machine learning algorithms, e.g. t-distributed stochastic neighbor embedding (t-SNE, \citealt{steinhardt2020}), are able to \editone{use a larger portion of the }SED information to classify galaxies, as opposed to only a select few bands. They have been shown to achieve this task quite well, boasting $\sim 80\%$ completeness and $<30\%$ contamination at $z<2$. \editone{One} major draw-back of these methods is that they are computationally-intensive and relatively complex (particularly those that employ non-parametric (flexible) models, e.g., \citealt{leja_sed_fitting}). Color selection, on the other hand, is convenient and accessible, especially as many catalogs now come with pre-computed rest-frame colors. 

Additionally, their performance hinges upon the availability of several photometric points, which is a luxury that many surveys with the next generation of telescopes (e.g. \emph{Euclid}, the \emph{Vera Rubin Observatory}, and the \emph{Nancy Grace Roman Space Telescope}) will not have. This will also be the case for \jwst\ observations in new fields, as these will have no existing ancillary data from other catalogs. Lastly, these alternative methods are model-dependent. SED-fitting depends on using galaxy templates that capture the variety of galaxies in the Universe and machine learning methods must be trained on large, representative samples of galaxies. Both of these are typically available at lower redshifts ($z <$ 3), and for several reasons (outlined above), we cannot assume that quiescent galaxies at low-z resemble those at high-z enough for these methods to be effective. When rest-frame colors are derived either directly from the observed photometry or \editone{using empirically-derived templates}, as we have done in this paper (Sections \ref{section:data} and \ref{section:extrapolation}), they are not subject to the underlying assumptions and biases associated with these model-dependent methods, hence are more representative of reality.  We argue that this will remain important through the \jwst\ era. 

Going forward, progress in the study of galaxies over cosmic time will likely continue to rely heavily on color selection techniques, both to distinguish between different galaxy populations and to identify targets for spectroscopy. \editone{Over the next few months, \emph{JWST} will provide a rapid and extremely efficient spectroscopic test\footnote{The cost of confirming quiescence at $z > 3$ comes out to $\sim \$250$ per object, assuming 15 hours of integration with Keck/MOSFIRE. With \emph{JWST}/NIRSPEC, this will be reduced to $\sim 1$ hour and $\sim \$10$ per object (assuming a 5-year mission).} of quiescence and hence of the \sugi\ color selection method. In the meantime, }the next step is to standardize these independent observational results and bring them to a common scale by using a single color\editone{-color selection method} over a range of redshifts, and we have demonstrated that the \sugi\ diagram can do that efficiently. In a follow-up paper, we will explore the distribution of galaxy properties at a range of redshifts in color-color space, comparing the global properties of star-forming and quiescent galaxies (e.g. stellar mass, star formation rate, number density, IR luminosity, stellar age, and morphology) of \sugi\, as has been done for \uvj\ (e.g. \citealt{williams2010}, \citealt{leja2015}, \citealt{patel2012}, \citealt{price2014}, \citealt{straatman2014}, \citealt{spitler2014}, \citealt{forrest2016}, \citealt{martis2016}, \citealt{fang2018}, \citealt{belli2019}). 

\begin{acknowledgements}
This work benefited from the generous support of the George P. and Cynthia Woods Mitchell Institute for Fundamental Physics and Astronomy at Texas A\&M University. This material is based upon work supported by the National Science Foundation under Grants AST-2009632 and AST-2009442. CMSS acknowledges support from Research Foundation— Flanders (FWO) through Fellowship 12ZC120N.

Based on observations taken by the 3D-HST Treasury Program (HST-GO-12177 and HST-GO-12328) with the NASA/ESA Hubble Space Telescope, which is operated by the Association of Universities for Research in Astronomy, Inc., under NASA contract NAS5-26555 and data products from observations made with ESO Telescopes at the La Silla Paranal Observatory under ESO program ID 179.A-2005 and on data products produced by TERAPIX and the Cambridge Astronomy Survey Unit on behalf of the UltraVISTA consortium. This research made extensive use of NASA’s Astrophysics Data System for bibliographic information.

JAD would like to thank Gabe Brammer for help with troubleshooting and running his SED-fitting code \texttt{eazy-py}, Justin Spilker and Rob Kennicutt for insightful discussions, and Taylor Hutchison and Jonathan Cohn for their help and support throughout the preparation of this manuscript. We are grateful to the anonymous referee whose feedback resulted in extremely valuable additions to the paper. 
\end{acknowledgements}

\section*{Data Availability}
The synthetic \sugi\ filters are available on Github\footnote{ \url{https://github.com/jacqdanso/synthetic-ugi-filters}}. Other data products can be provided upon request to the corresponding author.

\software{\texttt{Astropy} (\citealt{astropy2018}), 
\eazypy\ (\citealt{brammer2021}), 
\texttt{Jupyter} (\citealt{Kluyver2016jupyter}), 
\texttt{matplotlib} (\citealt{Hunter2007}), 
\texttt{numpy} (\citealt{harris2020array}), 
\texttt{pandas} (\citealt{pandas}), 
\texttt{seaborn} (\citealt{seaborn}),
\texttt{scipy} (\citealt{2020SciPy-NMeth}), 
\texttt{statsmodels} (\citealt{statsmodels})}
\newpage

\appendix
\section{Estimating Noise and Creating a sample of Dusty Star-Forming Galaxies in JAGUAR} \label{section:appendix}
\citet{williams2018} generated number counts of star-forming galaxies at $z>4$ in JAGUAR using UV luminosity functions, rather than stellar mass functions (Section \ref{subsection:simulations}). Dusty star-forming galaxies, however, tend to be underrepresented in UV-selected samples. Observations suggest that they comprise $\sim$ $50-60$\% of the log$_{10}$ (M$_*$/$M_\odot$) $> 10.5$ population at $z = 2-3$ \citep{martis2016},  and $\sim$ $40-50$\% at $z = 3-4$ (\citealt{spitler2014}, \citealt{martis2019}). JAGUAR features $\sim$ 20\% at these redshifts, a factor of at least 2 lower than what we expect from observations. This becomes problematic for evaluating completeness and contamination fractions of color-selected quiescent samples, as dusty star-forming galaxies are their primary contaminants. 

To resolve this, we artificially introduce more dusty star-forming galaxies into the JAGUAR catalogs by doing the following. For a subsample of galaxies with A$\rm _V \geq 1$ and $\log(\rm sSFR/yr)$ $\geq -9.5$, we determined the UV extinction at 1500 $\AA$ as a function of stellar mass using the following relation from \citet{panella2009} for log(M$_*$/M$_\odot$) $\geq 10.1$ star-forming galaxies at z $\approx 2$ , assuming that it does not evolve from z $= 2 - 6$.  
\begin{equation}
     A_{1500} = 4.07 \times \rm log(M_*) - 39.32
\end{equation}
For the low mass (log(M$_*$/M$_\odot$) $\leq 10.1$), we assume  A$_{1500}$ = 1 mag (\citealt{salim2005}). We applied 1 mag of scatter to A$_{1500}$, using a lognormal distribution to prevent the perturbed extinction values from going negative. Using the UV extinction, we estimated the color excess for each galaxy using the reddening curve from \citep{calzetti2000}:
\begin{equation}
     E(B-V) = \frac{A_{1500}}{k(1500)}
\end{equation}
The extinction as a function of wavelength, A($\lambda$), can then be obtained from the color excess, and the reddened flux in each bandpass is:
\begin{equation}
    f_{\nu, \rm reddened} = 10^{\frac{2.5\rm log (f_\nu)+A(\lambda)}{2.5}}
\end{equation}

We add uncertainties to the \emph{HST} and \jwst\ photometry using publicly-available code made for the JAGUAR catalogs (\citealt{hainline2020}). For a user-specified survey depth, the code estimates noise directly from the photometry by determining the total flux in the smallest fixed aperture between r=0.16$\arcsec$ and r=0.64$\arcsec$ that matches the galaxy's half-light radius, estimated using the Sersic index of the galaxy. The uncertainty on the flux in each band is then given as the Poisson noise and the instrument read noise summed in quadrature. We determine the final noise estimate per band (for each galaxy) as a random draw from a Gaussian with a width set to the uncertainty on the flux.  We show the \uvj, \ugi, and \sugi\ diagrams for log(M$_*$/M$_\odot$) $\geq$ 8.9 galaxies at $4 < z < 6$ using the perturbed data in Figure \ref{fig:jaguar_cc}. The mass cutoff is based on the expected mass completeness limit at $z = 6$ for the Cosmic Evolution Early Release Science (CEERS; P.I.: S. L. Finkelstein) survey (\citealt{kauffmann2020}). %

The addition of noise preferentially affects low mass (log(M$_*$/M$_\odot$) $<$ 10) and faint galaxies, causing them to scatter in and out of the quiescent region. The completeness of the perturbed sample is lower by $\sim 10\%$ from the original at $z = 3 - 4$ and $\sim 20\%$ at $z = 4 - 6$. The contamination rates after the addition of noise are similar to that of the unperturbed sample. 

\section{Effect of Photometric Redshift Uncertainty on Contamination from Strong Emission Lines} \label{section:appendix_b}
In Section \ref{subsection:eelgs} and Figure \ref{fig:high_ew}, we presented an estimate of the fraction of EELG contaminants in the quiescent selection of each color selection using a sample of galaxies at low-z ($1.2 < z < 2.3$) and high-z ($4 < z < 6$).  We demonstrated that their fractions in the \sugi$-$selected quiescent samples are lower by a factor of $1.6 - 3.5$ than that of \uvj\ at these redshifts, where the fraction of star-forming galaxies with rest-frame EW$ > 500$ \AA\  is expected to be $\simeq 10 -50$\% \citep{boyett2022}. Underlying these estimates, however, is the assumption that we have precise and accurate photometric redshifts. For the \threedhst\ sample, the maximum photo-z scatter (based on spectroscopic redshifts up to z $\sim 6$) is 2.6\% (\citealt{skelton2014}), and that for JAGUAR is effectively 0\%, since it is simulated data. In reality, redshift uncertainties for ground- and space-based photometric surveys may be higher. 

In this section, we consider the effect of photometric redshift uncertainty. We do this by estimating the photometric redshift uncertainty required for the strongest emission lines to shift into the synthetic filters (that is, we translate the wavelength shift required for a given emission line to contaminate a specified synthetic filter into a 1-$\sigma$ uncertainty on the redshift). Photometric redshift scatter is typically reported as the normalized median absolute deviation:
\begin{equation}\label{equation:z_nmad}
    \sigma_{NMAD} = 1.48 \times median\Bigg{(}\frac{\Delta z}{1 + z_{true}}\Bigg{)} 
\end{equation}
where $z_{true}$ is often determined using spectroscopy. Given an emission line that has shifted into a photometric bandpass 
\begin{equation}
    \lambda_{new} = \lambda_{orig}(1 + z + \Delta z) 
\end{equation}
the photometric redshift scatter required for the line to contaminate a given filter is
\begin{equation}\label{equation:z_scatter}
    \sigma_z = 1.48 \times\Bigg{(}\frac{\lambda_{new} -1}{\lambda_{orig}(1+z)}\Bigg{)}
\end{equation}

We report our results for the four strongest emission lines in Table \ref{table:sigma_z}. For blended lines and doublets, we make conservative estimates by using the central wavelength of the emission line closest to the synthetic filter in question. Based on our calculations, the filter most likely to get contaminated is \sg\ with \hb+\oiii. Due to the proximity of these emission lines to the filter, small photo-z uncertainties ($\simeq 5-9\%$) can cause the lines to scatter into the \sg\ filter and artificially boost the $(\su-\sg)_0$ color. Is this enough to warrant concern? Medium-band photometric surveys such as ZFOURGE (\citealt{straatman2016}) and NMBS (\citealt{whitaker2011}) boast $1-2$\% photo-z uncertainties due to finer sampling of the SED. Deep NIR surveys such as UltraVISTA (\citealt{muzzin2013}) that include medium-band photometry from the aforementioned also report similar uncertainties up to $z \sim 2$. Typical photo-z uncertainties are on the order of 5\% (e.g. COSMOS2020, \citealt{weaver2022}). Similarly, photo-z uncertainties of $5-8\%$ are expected for the faintest objects (m$_{F200W} > $ 24 AB) in upcoming photometric surveys with \jwst\ such as CEERS (\citealt{kauffmann2020}). Since these estimates are well within our redshift uncertainty tolerance for the \sg\ filter, contamination from galaxies with strong emission lines will likely be small compared to \uvj, as we have demonstrated in Section \ref{subsection:eelgs} .

A reasonable critique of this conclusion is that photo-z uncertainties from ground-based surveys are underestimated, as they are usually evaluated using galaxies with spectroscopic redshifts. These are typically highly biased samples, comprising only $\sim 1\%$ of the catalog and tend to be 2$-$5 magnitudes brighter than the rest of the sample. Our counter is that for precisely this reason, EELGs will have some of the most precise redshifts in the sample, as they tend to be bright and the presence of emission lines makes it easier to obtain spectroscopic redshifts for them. Additionally, many SED fitting codes (including \eazypy) now include extreme emission line galaxy templates in order to account for the effects of emission lines on the observed photometry (see Section \ref{subsection:rf_colors}). One way to test the accuracy of the photometric redshift uncertainties from our sample in \threedhst\ is to look at the predicted change in magnitude caused by a contaminating emission line. Because \sg\ is much narrower than $V$ ($\Delta \lambda = 400~ \AA$ versus $991~ \AA$), a stronger magnitude change is introduced in \sg\ than in $V$ when contaminated by an emission line (Equation \ref{equation:flux_boost}). This means that if the photo-z uncertainty exceeds our estimated threshold, there should be more galaxies with moderate \hb+\oiii\ line strengths (EW $>  200~ \AA$) in the \sugi\ quiescent region than that for \uvj, since they will produce a strong magnitude change of (1.93 mag) in the former and a relatively weaker one in the latter (0.64 mag). What we see is the opposite, with the number of EW $>  200~ \AA$ galaxies in \sugi\ being a factor of $\simeq 1.4$ less than that in \uvj\ (92 as opposed to 130). This suggests that the photometric redshift estimates in \threedhst\ have an uncertainty of $< 5\%$ and lends credibility to using spectroscopic redshifts for estimating photometric redshift uncertainties. 

\begin{deluxetable*}{|c|c|c|c|c|c|c|}
\tablecaption{Photometric redshift uncertainties required for the strongest emission lines to contaminate the synthetic filters.\label{table:sigma_z}}
\tablehead{
\colhead{Synthetic Filter} & 
\colhead{Filter Edge (\AA)} & 
\colhead{Emission Line} & 
\colhead{Rest Wavelength\tablenotemark{a} (\AA)}& \colhead{Buffer\tablenotemark{b} (\AA)} & \colhead{$\sigma_z$\tablenotemark{c}} & 
\colhead{$\Delta z$ (at $z = 3$)}
}
\startdata
 \su & 3100 (R) & [\ion{O}{2}]\ & 3729.875 & 627 & 0.249 & $-0.996$\\
 \hline
 \sg & 4300 (L) & [\ion{O}{2}]\ & 3729.875 & 573 & 0.228 & 0.912\\
 \sg & 4700 (R) & H$\beta$ & 4862.68 & 163 & 0.048 & $-0.192$\\
  \sg & 4700 (R) & \oiii & 4960.295, 5008.24 & 307 & 0.091 & $-0.364$\\
 \hline
 \si & 7000 (L) & H$\beta$+\oiii & 4862.68, 4960.295, 5008.24 & 2993\tablenotemark{d} & 0.885 & 3.54\\
 \si & 7000 (L) & H$\alpha$+\nii & 6564.61, 6585.27 & 452 & 0.102 & 0.408\\
 \si & 8000 (R) & \ion{Si}{3} & 9068.6, 9530.6 & 1069 & 0.174 & $-0.698$\\
\enddata
\tablenotemark{a}{Vacuum rest-frame wavelengths obtained from \citet{momcheva2016} and the SDSS database \footnote{\url{http://classic.sdss.org/dr6/algorithms/linestable.html}}.}

\tablenotemark{b}{Wavelength separation between the central wavelength of the line and filter edge, used to determine $\lambda_{new}$ in Equation \ref{equation:z_scatter}.}

\tablenotemark{c}{Photometric redshift scatter, given by Equation \ref{equation:z_scatter}.}

\tablenotetext{d}{Conservative estimate using \hb $\lambda$4863.}
\end{deluxetable*}

\section{Demographics of quiescent selections and contaminants of each method}\label{section:appendix_c}

\edittwo{Here, we sketch out the types of galaxies that are missing in \sugi\ selection and those that erroneously enter the sample. In particular, we answer the following questions: what galaxies are we missing and why? How are they biased relative to those selected? What are the false positives and why are we selecting them? To do this, we looked at galaxies at $3 < z < 4$ in UltraVISTA, as the effects of extrapolation (Section \ref{section:extrapolation}) and contamination from galaxies with strong emission lines (Section \ref{subsection:eelgs}) start to become more important at those redshifts. We show the SEDs and stellar population parameters of these galaxies, derived by fitting their UV-far IR photometry with \texttt{MAGPHYS} (Section \ref{section:uvista}).} 

\begin{figure*}
\begin{center}
\includegraphics[trim={2cm 0cm 1cm 1cm},clip,scale=0.45]{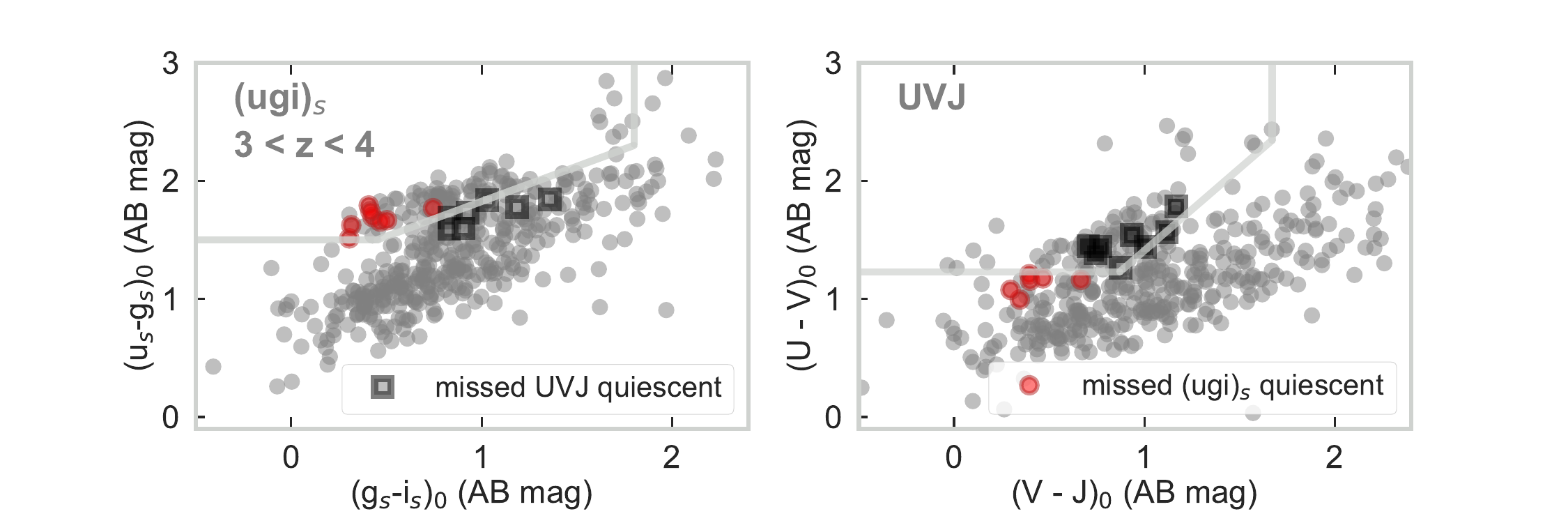}
\caption{\edittwo{UltraVISTA galaxies with log(M/M$_\odot$) $>$ 10.5 at $3 < z < 4$ selected as quiescent in \uvj\ that are not selected in \sugi\ and vice-versa. The \sugi-selected quiescent galaxies that are missed in \uvj\ lie in the region typically occupied by post-starbursts and young quiescent galaxies. This is confirmed by their SEDs and derived stellar population parameters in Figure \ref{fig:uvj_missed_seds}.}}
\label{fig:missed_objects_sugi_uvj}
\end{center}
\end{figure*}

\begin{figure*}
\begin{center}
\hspace*{-0.23cm} 
\includegraphics[trim={6.35cm 2cm 1cm 1cm},clip,scale=0.255]{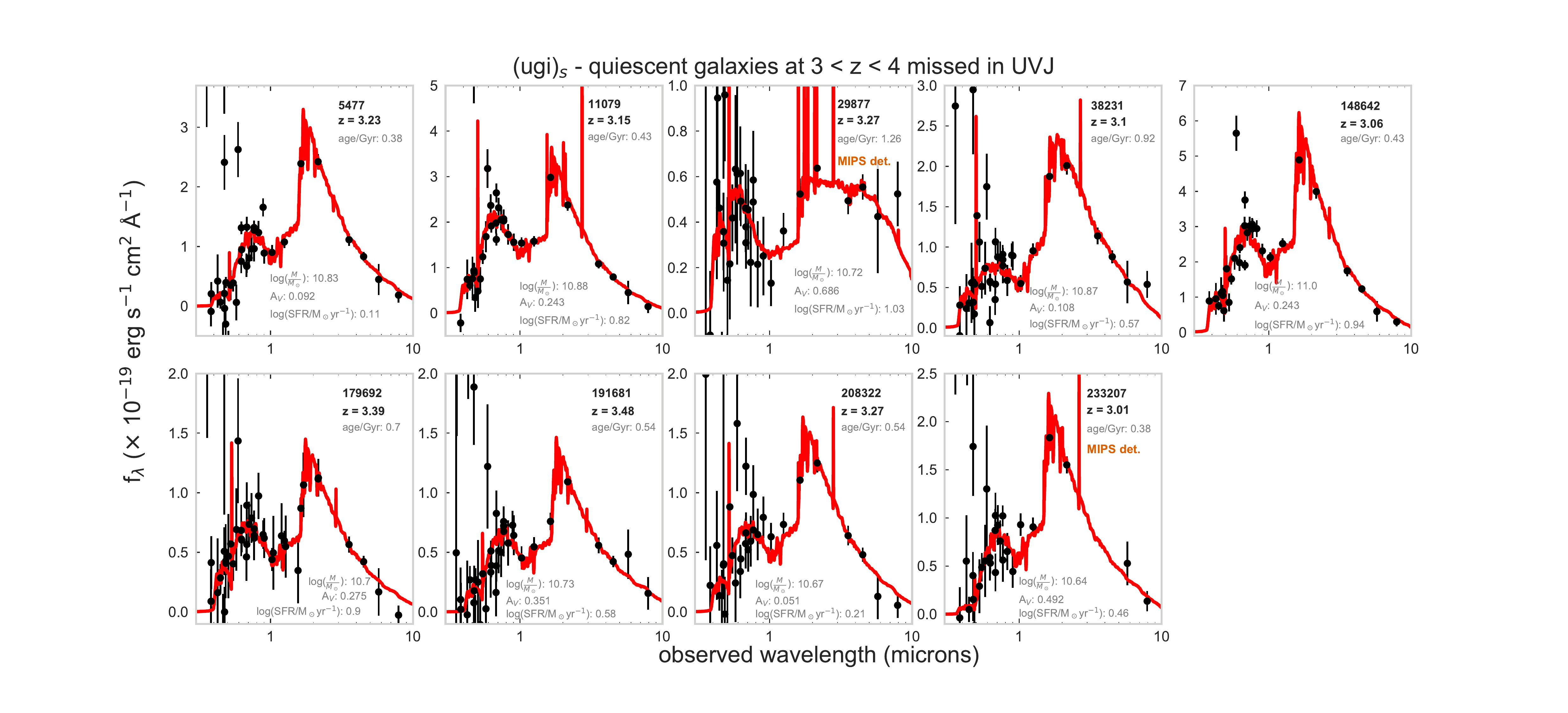}
\caption{\edittwo{Best-fit SEDs and derived stellar population parameters of \sugi-selected quiescent galaxies that are missed in \uvj\ selection at $3 < z < 4$ (red circles in Figure \ref{fig:missed_objects_sugi_uvj}). Eight out of nine are young (400 - 900 Myr) galaxies with relatively low star formation rates ($< 9$ M$_\odot$/yr) and dust extinction (A$_{\rm V}$ $<$ 0.5). The exception (ID=29877) was erroneously selected likely due to its high photometric redshift uncertainty ($\sigma_z$ $\sim 25\%$). Galaxies detected at SNR $>$ 3 in MIPS 24 $\mu m $ emission are labelled in orange.}}
\label{fig:uvj_missed_seds}
\end{center}
\end{figure*}

\begin{figure*}
\begin{center}
\includegraphics[trim={7.1cm 2cm 1cm 1cm},clip,scale=0.235]{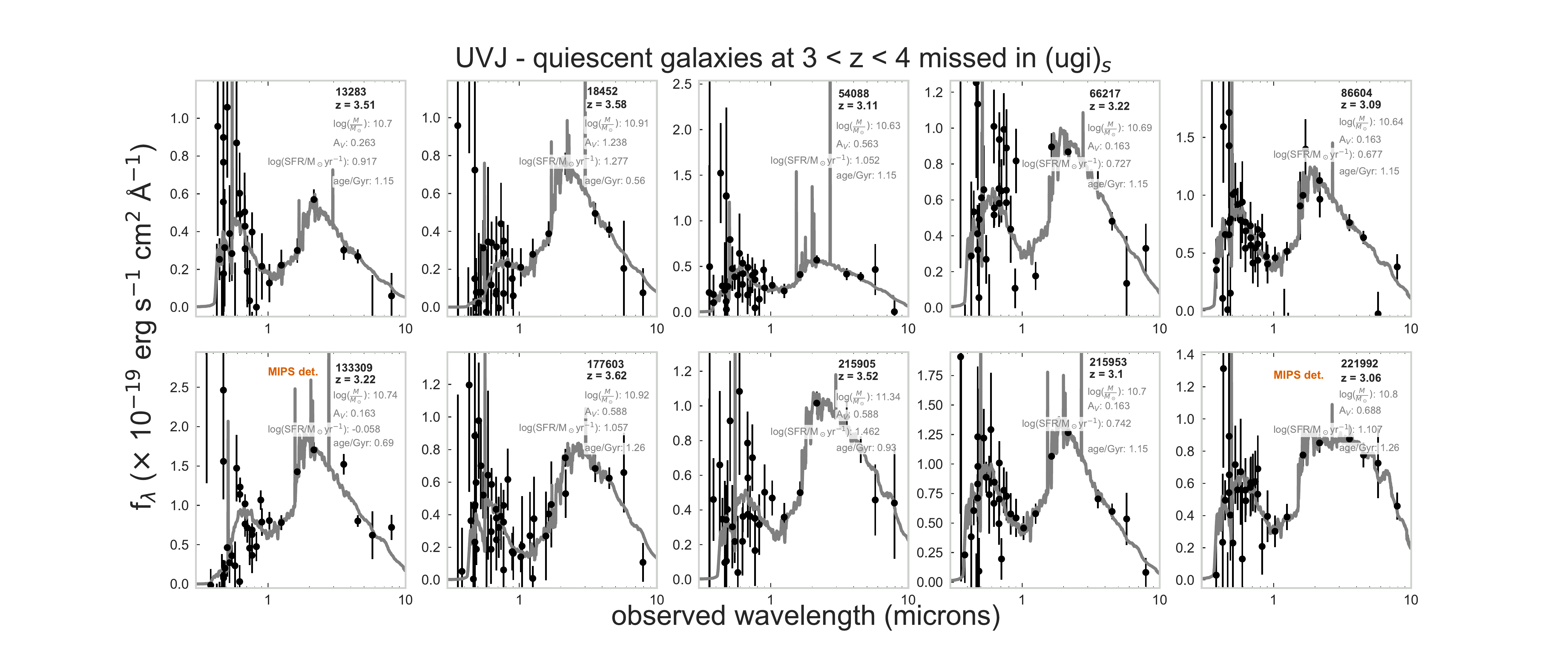}
\caption{\edittwo{Best-fit SEDs and derived stellar population parameters of \uvj-selected quiescent galaxies that are missed in \sugi\ selection at $3 < z < 4$  (gray squares in Figure \ref{fig:missed_objects_sugi_uvj}). These were erroneously selected likely due to emission line contamination.}}
\label{fig:ugs_missed_seds}
\end{center}
\end{figure*}

\begin{figure*}
\begin{center}
\includegraphics[trim={0cm 0.5cm 0cm 0cm},clip,scale=0.29]{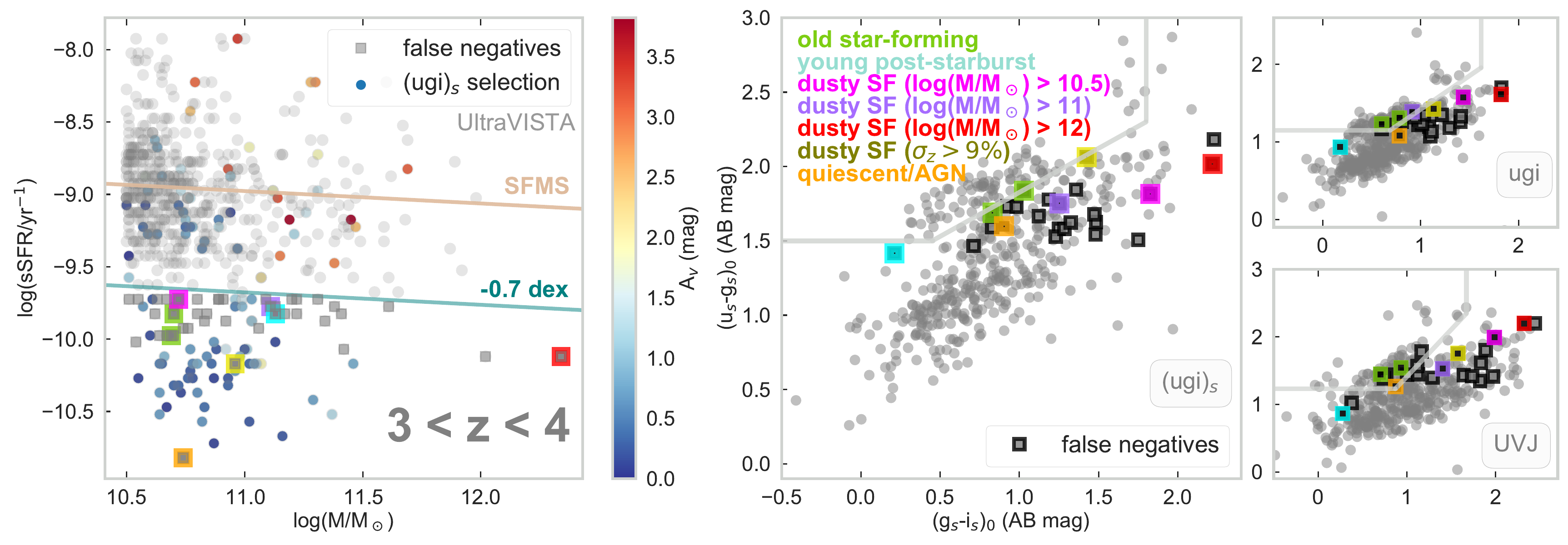}
\caption{\edittwo{{\bf Left:} Demographics of \sugi-selected quiescent galaxies at $3 < z < 4$ in UltraVISTA. Similar to Figure \ref{fig:demographics}, we have color-coded these galaxies by their best-fit dust extinction values from \texttt{MAGPHYS}. The gray squares show \emph{all} galaxies that were not selected as quiescent in \sugi. {\bf Right: } Locations of false negatives in \sugi, \ugi, and \uvj. 
The SEDs of the highlighted false negatives (colored squares) are shown in Figure \ref{fig:missed_seds}.}}
\label{fig:missed_demographics}
\end{center}
\end{figure*}

\begin{figure*}
\begin{center}
\hspace*{-1.3cm} 
\includegraphics[trim={2.1cm 0.5cm 0.5cm 0cm},clip,scale=0.275]{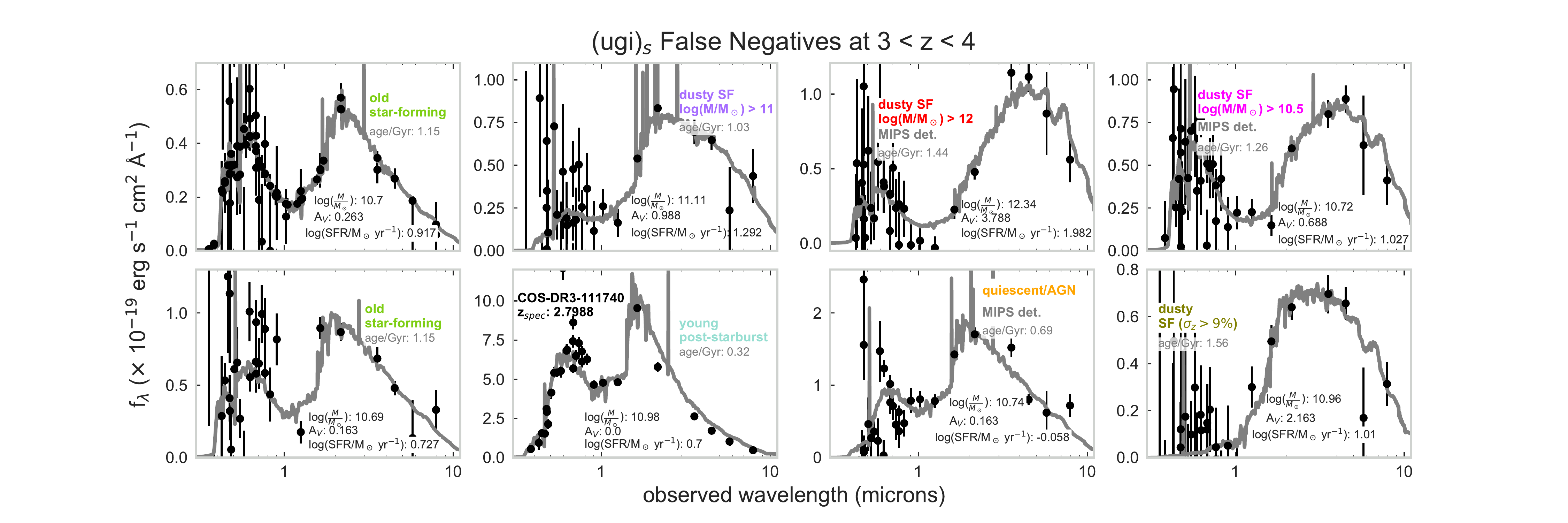}
\caption{\edittwo{SEDs and stellar population parameters of of low sSFR galaxies missed in \sugi\ selection at $3 < z < 4$. The SED and stellar population parameters of COS-DR3-111740 are based on fits to \emph{Keck}/MOSFIRE spectra in \cite{forrest2020_survey}. }}
\label{fig:missed_seds}
\end{center}
\end{figure*}

Our main conclusions are summarized as follows: 
\begin{itemize}
    \item \edittwo{\sugi\ selection captures recently-quenched galaxies at younger ages than \uvj\ and \ugi. This is evident from the color evolution tracks in Figures \ref{fig:sps_tracks} and \ref{fig:sps_tracks_comb}, with the $\tau = 100$ Myr line, also referred to as the "fast-quenching" track (e.g. \citealt{belli2019}), entering the \sugi\ quiescent region $\sim 250$ and 150 Myr earlier than in \uvj\ and \ugi, respectively. In Figure \ref{fig:missed_objects_sugi_uvj}, we show that the galaxies selected in \sugi\ that are missed in \uvj\ lie in the region typically occupied by post-starbursts and young quiescent galaxies. Their SEDs and derived fit parameters (Figure \ref{fig:uvj_missed_seds}) indicate that they are young (median age $\sim$ 0.5 Gyr) and have low dust extinction (median A$_{\rm V}$ $= 0.24$) and star formation rates (median $\sim 3.8 M_{\odot}/\rm yr$). Due to their central wavelengths, the \su and \sg\ filters are more sensitive to the Balmer break, which is strongest for galaxies that are 300 $-$ 500 Myr old (\citealt{d'eugenio2021}). The $U-V$ color on the other hand, is more sensitive to the 4000 \AA\ break, which becomes stronger than the Balmer break for populations that are $\sim 1$ Gyr old. }

    \item \edittwo{The galaxies missed in \sugi\ selection are generally not quiescent. They span a larger range in age, dust extinction, and star formation rate than the above. However, they fall into two broad categories: 1) young ($< 0.5$ Gyr) star-forming galaxies with low to moderate dust extinction (A$_{\rm V} < 0.5$) and 2) old ($> 1$ Gyr) star-forming galaxies with varying levels of dust. Although these galaxies are not selected in \sugi\ a number of them are selected in \uvj\ (Figure \ref{fig:ugs_missed_seds}) and \ugi\ (Figures \ref{fig:missed_demographics} and \ref{fig:missed_seds}). This suggests that we have underestimated our completeness rates, particularly for \sugi\ in Section \ref{subsection:completeness_contamination}.}

    \item \edittwo{The contaminants (false positives) in \sugi\ are 1) old ($> 1$ Gyr) galaxies with moderate (A$_{\rm V} = 0.5  - 0.8$) dust extinction and little to no emission lines. 2) Those with strong emission lines (which are few) tend to be  galaxies with large ($\sim 10-20\%$) photo-z uncertainties. 3) A handful of very young ($< 600$ Myr) dusty (A$_{\rm V} >$ 1) star forming galaxies also contaminate the \sugi\ quiescent region. 4) Galaxies with high dust extinction (A$_{\rm V} > 2$) tend to cluster at (\su$-$\sg) $>$ 2.4, therefore, they may be culled using this criterion. 5) Catastrophic outliers ($\sigma_{z} > 50\%$) contaminate the quiescent region in all three methods. }
\end{itemize}

\bibstyle{aasjournal}
\bibliography{paper_merged.bib} 

\end{document}